\documentclass[12pt]{article} 
\usepackage{graphicx,subfigure,latexsym,amssymb}

\newcommand{\be}{\begin{equation}}
\newcommand{\ee}{\end{equation}}
\newcommand{\bear}{\begin{eqnarray}}
\newcommand{\eear}{\end{eqnarray}}
\newcommand{\ba}{\begin{array}}
\newcommand{\ea}{\end{array}}

\def\del{{\partial}}
\def\({\left(}
\def\){\right)}
\def\ka{\kappa}

\def\omg{\omega}

\def\Dlt{\Delta}
\def\sig{\sigma}

\begin{document}

\begin{titlepage}
\vfill
\begin{flushright}
{\normalsize RBRC-1018}\\
\end{flushright}

\vfill
\begin{center}
{\Large\bf  Out-of-Equilibrium Chiral Magnetic Effect  \\at Strong Coupling }

\vskip 0.3in
Shu Lin$^{1}$\footnote{e-mail: {\tt  slin@quark.phy.bnl.gov}},
Ho-Ung Yee$^{1,2}$\footnote{e-mail:
{\tt hyee@uic.edu}}
\vskip 0.3in

{\it $^{1}$RIKEN-BNL Research Center, Brookhaven National Laboratory,} \\
{\it Upton, New York 11973-5000 }\\[0.15in]
 {\it $^{2}$Department of Physics, University of Illinois,} \\
{\it Chicago, Illinois 60607 }\\[0.3in]


\end{center}

\vfill

\begin{abstract}
We study the charge transports originating from triangle anomaly in out-of-equilibrium conditions 
in the framework of AdS/CFT correspondence at strong coupling, to gain useful insights on possible charge separation effects that may happen in the very early stages of heavy-ion collisions.
We first construct a gravity background of a homogeneous mass shell with a finite (axial) charge density gravitationally collapsing to a charged blackhole, which serves as a dual model for out-of-equilibrium
charged plasma undergoing thermalization. We find that a finite charge density in the plasma slows down the thermalization. We then study the out-of-equilibrium properties of Chiral Magnetic Effect and Chiral Magnetic Wave in this background.  As the medium thermalizes, the magnitude of chiral magnetic conductivity and the response time delay grow. We find a dynamical peak in the spectral function of retarded current correlator, which we identify as an out-of-equilibrium chiral magnetic wave. The group velocity of the out-of-equilibrium chiral magnetic wave is shown to receive a dominant contribution from a non-equilibrium effect, making the wave moving much faster than in the equilibrium, which may enhance the charge transports via triangle anomaly in the early stage of heavy-ion collisions.
\end{abstract}

\vfill

\end{titlepage}
\setcounter{footnote}{0}

\baselineskip 18pt \pagebreak
\renewcommand{\thepage}{\arabic{page}}
\pagebreak

\section{Introduction}

Heavy-ion collisions create an interesting new state of matter, quark-gluon plasma of QCD, where confinement is effectively lost
due to high temperature above the QCD cross-over line.
Although microscopic QCD degrees of freedom of quarks and gluons are expected to be liberated in this environment, there are many experimental and theoretical
indications that the quark-gluon plasma created in the experiments are strongly coupled, which makes them behaving as nearly perfect liquids with small viscosity \cite{Shuryak:2008eq}.
Hydrodynamics has been a powerful tool to describe the long wavelength dynamics of the system without knowing much about the microscopic details of the theory except a few transport coefficients. However, going beyond the hydrodynamic regime meets a serious computational challenge of dealing with 
strongly coupled system of QCD matter. The AdS/CFT correspondence based on a large $N_c$ expansion and strong t'Hooft coupling can be a useful tool
to study such strongly coupled QCD dynamics.

Another approach to circumventing difficulties of strongly coupled dynamics is to use symmetries of the theory
and look for interesting observables that are protected by them.
QCD with (approximately) massless quarks has a chiral flavor symmetry $SU(N_f)_L\times SU(N_f)_R\times U(1)_V\times U(1)_A$, where the last axial symmetry $U(1)_A$ is quantum mechanically violated via triangle anomaly, and it is not a true symmetry. The gluonic contributions in the plasma to the anomalous violation of axial symmetry
happen via thermal sphaleron transitions, whose rate in current estimate is about $\Gamma_{\rm sph}\approx  30\alpha_s^5 T^4 \approx 0.12 \alpha_s^5\,{\rm GeV}^4$ with $T=250$ MeV \cite{Moore:2010jd}.
This determines the relaxation time scale of axial charges via fluctuation-dissipation relation as
\be
\tau_{\rm R} ={2\chi T\over (2N_F)^2 \Gamma_{\rm sph}}\approx {3.1\times 10^{-3}\over \alpha_s^5}\,{\rm fm}\,,
\ee where $\chi\approx 1.0\,T^2\approx 0.06 \,{\rm GeV}^2$ is the charge susceptibility at $T=250$ MeV \cite{Hegde:2008rm}, and $N_F=2$. This gives $\tau_{\rm R}\approx 10$ fm for $\alpha_s=0.2$, and one could marginally neglect it in heavy-ion experiments with typical lifetime of the plasma being 10 fm \footnote{The above relaxation time formula is highly sensitive to $\alpha_s$: for example, $\alpha_s=0.3$ reduces it to $\tau_{\rm R}=1.3$ fm. However, our main purpose of this work is about early time of $t\lesssim 1$ fm, so we can still neglect 
sphaleron relaxation for our work. We also stress that the main sources of the axial charge in such early time should be color electric/magnetic fields from glasma \cite{Kharzeev:2001ev,Lappi:2006fp}.}.
Another (more formal) aspect of these gluonic contributions is that they are sub-leading in large $N_c$ limit, and
would not appear, for example, in the AdS/CFT-based models at leading order.  

Instead of having gluonic fields, a flavor gauge field such as electromagnetic field can give rise to the same type of triangle
anomaly of the axial symmetry,
\be
\partial_\mu J_A^\mu={e^2 N_c \over 2\pi^2}\left(\sum_F q_F^2\right) \vec E\cdot\vec B\,,
\ee where $q_F$ is the charge of the quark flavor $F$.
Non-renormalization of this relation under radiative corrections is a rare example where a violation of a symmetry
can give us strong constraints on the predictions of the theory. In the low energy regime of chiral perturbation theory,
the gauged Wess-Zumino-Witten action accounts for all essential physics consequences of the triangle anomaly.
However, possible new transport phenomena originating from triangle anomaly in finite temperature or density
phases of QCD are less explored and have attracted much recent interests from both theorists and experimentalists.  One such phenomenon, the Chiral Magnetic Effect (CME) \cite{Kharzeev:2007tn,Kharzeev:2007jp,Fukushima:2008xe,Son:2004tq,Metlitski:2005pr}, states that in the presence of a magnetic field $\vec B$, a vector (axial) current will be induced by a non-zero axial (vector) chemical potential,
\be
\vec J_{V,A}={eN_c\over 2\pi^2}\mu_{A,V}\vec B\,.
\ee
The CME has been confirmed in both weak coupling \cite{Kharzeev:2009pj,Hong:2010hi,Hou:2011ze,Loganayagam:2012pz,Son:2012wh,Stephanov:2012ki,Zahed:2012yu,Chen:2012ca} and strong coupling frameworks \cite{Yee:2009vw,Rebhan:2009vc,Gorsky:2010xu,Gynther:2010ed,Kalaydzhyan:2011vx,Hoyos:2011us}. It has also been derived from the hydrodynamics \cite{Son:2009tf,Neiman:2010zi} and effective action \cite{Lublinsky:2009wr,Sadofyev:2010is,Lin:2011aa,Nair:2011mk,Bhattacharya:2012zx,Kalaydzhyan:2012ut}.
The off-central heavy-ion collisions which accompany transient magnetic fields of strength as large as $eB\sim m_{\pi}^2$ are important places to look for possible signals of this effect \cite{Kharzeev:2007jp}, and there are experimental indications
which favor the existence of the signals that go along with the predictions from the Chiral Magnetic Effect \cite{Voloshin:2004vk,Abelev:2009ac,Selyuzhenkov:2011xq}.

The two versions of the Chiral Magnetic Effect lead to the existence of a new gapless sound-like propagating
mode of chiral charge densities in the hydrodynamic regime, coined as Chiral Magnetic Wave (CMW) \cite{Kharzeev:2010gd,Newman:2005hd}, which has the
dispersion relation,
\be
\omega=\mp v_\chi k-iD_L k^2 +\cdots\,,
\ee 
where the velocity $v_\chi$ is given by $v_\chi={eN_c B\over 4\pi^2 \chi}$, and $k$ is the momentum along the direction of the magnetic field. The longitudinal diffusion constant $D_L$ depends more on the dynamics of the theory. The sign in front of the first term that determines the direction of the wave propagation depends on the chirality of the charge fluctuations, so that a left-handed chiral charge fluctuation moves to the direction opposite to a right-handed chiral charge fluctuation. In off-central heavy-ion collisions, the charge transports 
via Chiral Magnetic Wave would induce a net electric quadrupole moment in the fireball \cite{Burnier:2011bf,Burnier:2012ae,Gorbar:2011ya}, which eventually leads to
a charge dependent elliptic flow of pions \cite{Burnier:2011bf,Burnier:2012ae}. Recent analysis from {\bf STAR} seems to support the prediction from the Chiral Magnetic Wave \cite{Wang:2012qs,Ke:2012qb}.
Both Chiral Magnetic Effect and Chiral Magnetic Wave above should be considered as long wavelength limit
of the charge transports originating from triangle anomaly in the equilibrium QCD plasma. 

In this work, we extend the previous studies in two important aspects: we study Chiral Magnetic Effect and Chiral Magnetic Wave in out-of-equilibrium conditions and in non-hydrodynamic regimes.
By out-of-equilibrium conditions, we mean that the plasma background in question is not thermalized and non-static either. By non-hydrodynamic regimes, we mean the frequency of the probe (in our case, it will be the magnetic field) is 
comparable or larger than the characteristic time scale of the plasma loosely set by the late-time temperature or effective collision rate.
Our motivation for considering out-of-equilibrium plasma is to study the charge transport originating from triangle anomaly in the early stages of plasma fireball created in heavy-ion collisions where the system is out-of-equilibrium and undergoes thermalization. This is well motivated since the magnetic field is larger at earlier times
and the charge transports via triangle anomaly may be significant in this out-of-equilibrium stage before local thermalization is achieved. Since the thermalization seems to happen relatively fast within 1 fm, 
how large the net effects coming from the out-of-equilibrium stage are is an important question to be addressed carefully. We hope our work lays a useful foundation to answer this question more quantitatively in the future.
The motivation for looking at non-hydrodynamic response to a magnetic field of high frequency, which was first
studied in \cite{Kharzeev:2009pj} at weak coupling and subsequently in \cite{Yee:2009vw} at strong coupling,
comes from the fact that the magnetic field created in heavy-ion collisions is highly time-dependent and transient.
When the frequency $\omega$ of the magnetic field is finite, the Chiral Magnetic Effect is generalized to be
\be
\vec J_V(\omega)=\sigma_\chi(\omega)\vec B(\omega)\,,
\ee 
with the frequency-dependent chiral magnetic conductivity $\sigma_\chi(\omega)$.
In the equilibrium QCD plasma, its zero frequency limit is constrained to reproduce the usual Chiral Magnetic Effect, so that 
\be
\sigma_\chi(\omega\to 0)={eN_c\over 2\pi^2}\mu_A \,,
\ee
whereas the finite frequency behavior depends on the microscopic dynamics of the theory.
In our analysis, we look at the same problem in out-of-equilibrium conditions.

We will study these problems in the framework of AdS/CFT correspondence, hoping to gain useful insights
on what would be the results at strong coupling \footnote{See Refs.\cite{Lin:2006rf,Lin:2008rw,Bhattacharyya:2009uu,Beuf:2009cx,Chesler:2009cy,Bantilan:2012vu,Balasubramanian:2010ce,CaronHuot:2011dr,Garfinkle:2011hm,Wu:2011yd,Galante:2012pv,Caceres:2012em,Baron:2012fv,Chesler:2012zk} for previous works on out-of-equilibrium situations in AdS/CFT correspondence.}. In AdS/CFT, global symmetries such as vector/axial symmetries
appear as 5-dimensional gauge fields residing in the holographic 5 dimensional AdS space. 
For our purposes, we can focus on simply $U(1)_V\times U(1)_A$, and the triangle anomaly
manifests itself as a 5 dimensional Chern-Simons term,
so that the minimal set-up of our holographic model is the 5 dimensional Einstein-Maxwell-Chern-Simons theory
with $U(1)_V\times U(1)_A$ gauge fields,
\bear
\left(16\pi G_5\right){\cal L}&=&R+12-{1\over 2}\left(F_{V}\right)_{MN} \left(F_{V}\right)^{MN}
-{1\over 2}\left(F_A\right)_{MN}\left(F_A\right)^{MN}\\
&+&{\kappa\over 2\sqrt{-g_5}}\epsilon^{MNPQR}\left(3 \left(A_A\right)_M\left(F_V\right)_{NP}\left(F_V\right)_{QR}
-\left(A_A\right)_M\left(F_A\right)_{NP}\left(F_A\right)_{QR}\right)\,.\nonumber
\eear
The coefficient $\kappa$ of the Chern-Simons terms should be chosen as
\be
\kappa=-{2 G_5 N_c\over 3\pi}\,,\label{kappa}
\ee
to reproduce the correct triangle anomaly with a single massless Dirac quark flavor whose electromagnetic charge is set to $e$. Our epsilon symbol is purely numerical with the convention $\epsilon^{zt123}=1$\footnote{Note that our definition of epsilon tensor differs by a sign from that in Ref.\cite{Yee:2009vw} because our radial coordinate $z$ is related to the coordinate $r$ in Ref.\cite{Yee:2009vw} by $z={1\over r}$, which is a parity odd transformation. Thus we have an overall plus sign for the Chern-Simons term.}. Note that our vector gauge field $A_V$ is defined to be dual to the vector current without $e$, so that
the electromagnetic current is the $e$ times the vector current obtained from $A_V$. Similarly, an electromagnetic background field will act as a source for the vector current with the coupling $e$, so that the boundary value of $A_V$ will be $e$ times the electromagnetic background field. 
  The generalization to multi-flavor quarks with different electromagnetic charges is straightforward with a few rescalings of parameters. The 5 dimensional Newton's constant $G_5$ in our model can be fixed by
considering the equation of state of the blackhole solution that describes finite temperature QCD plasma at high temperatures
\be
{\varepsilon\over T^4}={3\pi^3\over 16 G_5}\,,
\ee
and comparing this with the lattice result for $T\gg T_c$ \cite{Philipsen:2012nu},
\be
{\varepsilon\over T^4}\approx 13\quad ({\rm lattice})\,,
\ee
which gives $G_5\approx 0.45$.

We will first construct a background geometry of our theory for out-of-equilibrium conditions, generalizing
the falling mass shell geometry used in \cite{Lin:2008rw}, now including a finite axial charge density on the shell to discuss the
Chiral Magnetic Effect~\footnote{See Refs.\cite{Erdmenger:2012xu,Baier:2012tc,Steineder:2012si,Steineder:2013ana} for works on similar geometries with zero charge density.}. 
Independently to the Chiral Magnetic Effect, our inclusion of a finite charge is also motivated by the fact that the created fireball in heavy-ion collisions carries a finite vector chemical potential due to baryon stopping, and we would like to understand its effect on thermalization\footnote{The effects of vector and axial charge density will be the same in our model.}. 
Our model implicitly assumes the creation of axial charge fluctuation very early in the collision history, probably by color electric and magnetic fields in the glasma phase \cite{Kharzeev:2001ev,Lappi:2006fp}.
The Chern-Simons terms do not play a role in constructing the background solution, and one can patch
the known AdS-Reisner-Nordstrom blackhole solution in the UV region above the shell with the pure AdS solution in the IR region below the shell. Assuming the conformal energy-momentum tensor on the shell,
the Israel junction conditions \cite{Israel:1966rt} result in a simple equation for the time-trajectory of the shell, which we solve numerically. The initial position of the shell at time zero measures the typical virtuality scale of the initial out-of-equilibrium plasma, and it is natural to set it to be equal to the saturation scale $Q_s\sim 0.87$ GeV for {\bf RHIC} and 
$Q_s\sim 1.23 $ GeV for {\bf LHC}\footnote{Our values are based on the fit formula $Q_s^2\approx0.26 A^{1\over 3}\left(x\over 0.001\right)^{-0.3}$ in 
Ref.\cite{Kowalski:2007rw} with $x=0.01$ for {\bf RHIC} and $x=0.001$ for {\bf LHC}.}. As for the late-time equilibrium temperature, we will put $T=300$ MeV for {\bf RHIC} and $T=400$ MeV for {\bf LHC} 
as exemplary values. With these two scales fixed, the solution is unique given the (axial) charge density
(or equivalently, the late-time equilibrium value of the (axial) chemical potential, $\mu_A^{eq}$).
We will present our results for the values of $\mu_A^{eq}=50, 100, 200$ MeV.

In these new backgrounds, we study the charge transports originating from triangle anomaly via the 5 dimensional Chern-Simons terms.
We first study the frequency-dependent chiral magnetic conductivity $\sigma_\chi(\omega)$ in an approximation that the mass shell at a given time is nearly static compared to the time scale of the probe (quasi-static approximation) \cite{Lin:2008rw}. This brings us some constraints on the validity of our results, and
the precise region of validity will be discussed in detail. The time-trajectory of the mass shell then allows us to
find the time evolution of the chiral magnetic conductivity, $\sigma_\chi(\omega,t)$, in the quasi-static approximation. Going beyond the quasi-static approximation will be an interesting future direction to pursue.

We next study the time evolution of the Chiral Magnetic Wave dispersion relation in the neutral falling mass shell geometry, again in the quasi-static approximation. 
In this case, we assume a homogeneous, static background magnetic field which solves the equations of motion trivially, and we are interested in how chiral charge fluctuations behave under this condition, treating them as linearized small fluctuations.
We are interested in not only the low momentum regime, but also in non-hydrodynamic regime of finite spatial momenta, envisioning that the relevant charge fluctuations
in the heavy-ion collisions may be highly inhomogeneous in the transverse plane. For such large frequency-momentum regime, the quasi-static approximation is also better justified.
We will look for wave-like excitations in the spectral function below the lightcone, which is the region we expect to see chiral magnetic wave.


\section{Falling mass shell in AdS with finite charge density}

In this section, we will construct a gravitationally collapsing mass shell geometry in asymptotic $AdS_5$ space
with a 3-dimensional translational symmetry, generalizing previous works by including a finite axial charge (below we will simply call charge) density on the shell. This geometry is a toy model for a spatially homogeneous, out-of-equilibrium, charged plasma which undergoes thermalization. The late-time asymptotic solution which is dual to a thermally equilibrated charged plasma will be the known charged blackhole solution in $AdS_5$.
From our action density
\bear\label{ad}
\left(16\pi G_5\right){\cal L}&=&R+12-{1\over 2}\left(F_{V}\right)_{MN} \left(F_{V}\right)^{MN}
-{1\over 2}\left(F_A\right)_{MN}\left(F_A\right)^{MN}\\
&+&{\kappa\over 2\sqrt{-g_5}}\epsilon^{MNPQR}\left(3 \left(A_A\right)_M\left(F_V\right)_{NP}\left(F_V\right)_{QR}
-\left(A_A\right)_M\left(F_A\right)_{NP}\left(F_A\right)_{QR}\right)\,,\nonumber
\eear
with $\kappa=-{2G_5 N_c\over 3\pi}$, the equations of motion read as
\bear
&&R_{MN}+\left(4+{1\over 6}\left(F_V\right)^2+{1\over 6}\left(F_A\right)^2\right) g_{MN}
-\left(F_V\right)_{PM}\left(F_V\right)^P_{\,\,\,\,N}
-\left(F_A\right)_{PM}\left(F_A\right)^P_{\,\,\,\,N}=0\,,\nonumber\\
&& \partial_N\left(\sqrt{-g_5}\left(F_A\right)^{MN}\right)-{3\kappa\over 4}\epsilon^{MNPQR}\left(\left(F_V\right)_{NP}\left(F_V\right)_{QR}+\left(F_A\right)_{NP}\left(F_A\right)_{QR}\right)=0\,,\nonumber\\
&& \partial_N\left(\sqrt{-g_5}\left(F_V\right)^{MN}\right)-{3\kappa\over 2}\epsilon^{MNPQR}\left(F_A\right)_{NP}\left(F_V\right)_{QR}=0\,.\label{mastereom}
\eear
The model has an exact charged black-hole solution which is spatially homogeneous (AdS-Reisner-Nordstrom (AdS-RN) solution),
\be
ds^2={dz^2\over f(z) z^2}-{f(z)\over z^2}dt^2 + {\left(d\vec x\right)^2\over z^2}\,,\quad A_A=-Qz^2 dt\,,\quad A_V=0\,,
\ee
where 
\be
f(z)=1-mz^4+{2Q^2\over 3} z^6\,,
\ee
and $z_H$ is the location of the blackhole horizon obtained by solving $f(z_H)=0$.
The parameters $(m,Q)$ are related to the temperature and (axial) chemical potential $(T,\mu_A)$ by
\be
T=-{f'(z_H)\over 4\pi}\,,\quad \mu_A=z_H^2 Q\,.\label{Tmu}
\ee
The model also has the pure $AdS_5$ solution,
\be
ds^2={dz^2\over z^2}-{dt^2\over z^2}+{\left(d\vec x\right)^2\over z^2}\,,\quad A_V=A_A=0\,,
\ee 
corresponding to the vacuum of the model.

We will consider a thin, spatially homogeneous mass shell with a finite charge density collapsing from the UV region of small $z$ to the IR region of large $z$ under its own gravity.
Following \cite{Lin:2008rw}, we approximate the thickness of the shell to be infinitesimally small, and the geometry will be constructed by joining the AdS-RN solution above the shell in the UV region with the pure AdS solution below the shell, across the space-time trajectory of the thin mass shell which should be obtained by solving the appropriate Israel junction conditions \cite{Israel:1966rt}. In general, the coordinates $(z,t,\vec x)$ appearing in the AdS-RN solution above the shell should not be identified with the $(z,t,\vec x)$ in the pure AdS below the shell, and one should specify proper relations between them. One of the junction conditions is the continuity of the metric across the shell, so that the two metrics evaluated on the 1+3 dimensional world volume $\Sigma$ of the shell should be equal.
A part of this condition can easily be satisfied for the 3-dimensional spatial directions parametrized by $\vec x$, by
identifying $(z,\vec x)$ in the AdS-RN and $(z,\vec x)$ in the pure AdS across the shell, so that the metric part
${1\over z^2}{\left(d\vec x\right)^2}$ in both solutions match across the shell. After this, the time coordinates in the upper region (above the shell) and in the lower region (below the shell) are in general different, so we call them $t_U$ and $t_L$ respectively.
It is convenient to introduce a 1+3 dimensional world-volume coordinate $(\tau,\vec x)$ on the mass shell, and the induced metric on the shell can always be put into the form
\be
ds^2_\Sigma ={-d\tau^2 +\left(d\vec x\right)^2\over \left(z(\tau)\right)^2}\,,\label{WSmetric}
\ee
by reparameterizing $\tau$ and some function $z(\tau)$. By identifying $\vec x$ on $\Sigma$ with $\vec x$ in the background, $z(\tau)$ is clearly the position of the shell in the $z$ coordinate at time $\tau$. The remaining relations between $t_U$, $t_L$, and $\tau$, and the mass shell trajectory $z(\tau)$ (equivalently, $z(t_U)$ and $z(t_L)$) should be found by solving the junction conditions.

The continuity of the metric across the shell implies that the time component of the metric should match.
Writing the trajectory of the shell in the AdS-RN coordinates $(t_U,z)$ parametrized by the world sheet time $\tau$,
\be
(t_U,z)=\left(t_U(\tau),z(\tau)\right)\,,
\ee
and comparing the induced metric on the shell from the AdS-RN and (\ref{WSmetric}), one obtains
\be
f\left(z(\tau)\right)\dot t_U^2(\tau)-{\dot z^2(\tau)\over f\left(z(\tau)\right)}=1\,,\label{eq1}
\ee
where $\cdot\equiv{d\over d\tau}$. Similarly, the same trajectory in the pure AdS coordinates
\be
(t_L,z)=\left(t_L(\tau),z(\tau)\right)\,,
\ee
should satisfy the condition
\be
\dot t_L^2(\tau)-\dot z^2(\tau)=1\,.\label{eq2}
\ee
The (\ref{eq1}) and (\ref{eq2}) implicitly give the relation between $t_U$ and $t_L$ once the trajectory $z(\tau)$ is found. The last ingredient to determine the solution is the Israel junction condition\footnote{One can show that the extra terms from the gauge field in the Einstein equation does not modify the junction condition for the metric, as the field strengths do not contain $\delta$-function singularity. Some derivatives of the field strength such as $\partial_z F_{tz}$ are $\delta$-function singular, and they modify the junction condition for the gauge field coming from the Maxwell(-Chern-Simons) equations, which is nothing but the Gauss's law across the thin shell. We will not need to consider this in our work.} 
\be
\left[K_{ij}-\gamma_{ij}K\right]=-8\pi G_5 S_{ij}\,,
\ee
where $[A]\equiv A_L-A_U$ and $S_{ij}$ is the energy-momentum on the shell,
\be
S_{ij}={-2\over\sqrt{-\gamma}}{\delta\left(\sqrt{-\gamma}{\cal L}_{shell}\right)\over\delta\gamma^{ij}}\,,
\ee
and $\gamma_{ij}$ is the induced metric on the shell with respect to the shell coordinate $\xi^i$. The $K_{ij}^{U,L}$ are the extrinsic curvatures evaluated on
the shell from the upper region (AdS-RN metric) and the lower region (pure AdS) respectively,
\be
K_{ij}={\partial x^\alpha\over\partial\xi^i}{\partial x^\beta\over\partial \xi^j} \nabla_\alpha n_\beta=-n_\alpha\left({\partial^2 x^\alpha\over\partial \xi^i\partial\xi^j}+\Gamma^\alpha_{\beta\gamma} {\partial x^\beta\over\partial\xi^i}{\partial x^\gamma\over\partial\xi^j}\right)\,,
\ee
with the unit normal $n^\mu$ to the surface $\Sigma$ pointing to the direction of increasing $z$ (that is, out-going from the upper region of small $z$ to the lower region of large $z$).
Explicitly, $n^\mu$ in the upper and lower coordinates are given by
\bear
n_U&=&\left({z\dot z\over f(z)}\right){\partial\over\partial t}+\left(zf(z)\dot t\right){\partial\over\partial z}\,,\nonumber\\
n_L&=&\left(z\dot z\right){\partial\over\partial t}+\left(z\dot t\right){\partial\over\partial z}\,,\label{nul}
\eear
where all quantities are evaluated on the shell.

A straightforward computation gives the non-vanishing components as
\bear
K^U_{\tau\tau}&=&-{\dot t_U\over z}\left({f\left(f'+2\ddot z\right)\over 2\left(f+\dot z^2\right)}-{f\over z}\right)\,,\quad
K^U_{ij}=-{\dot t_U f\over z^2} \delta_{ij}\,,i,j=1,2,3\,,\nonumber\\
K^L_{\tau\tau}&=&-{\dot t_L\over z}\left({2\ddot z\over 2\left(1+\dot z^2\right)}-{1\over z}\right)\,,\quad
K^L_{ij}=-{\dot t_L \over z^2} \delta_{ij}\,,i,j=1,2,3\,.
\eear
where $'\equiv {d\over dz}$. To proceed further, we assume that the energy-momentum on the shell has the conformal form,
\be
S_{ij}=4p(z)u_i u_j +\gamma_{ij} p(z)\,,\quad u_i=\left({1\over z},0,0,0\right)\,,
\ee
with the pressure $p(z)$ to be determined, and the junction condition becomes after some manipulations,
\bear
\dot t_L-f \dot t_U=8\pi G_5 p(z)\,,\quad
\dot t_L{z\ddot z\over(1+\dot z^2)}-\dot t_U{zf\left({f'\over 2}+\ddot z\right)\over (f+\dot z^2)}=4\cdot8\pi G_5 p(z)\,.
\eear
Removing $p(z)$ from the above equations and using
\be
\dot t_L=\sqrt{1+\dot z^2}\,,\quad \dot t_U={\sqrt{f+\dot z^2}\over f}\,,
\ee
from (\ref{eq1}) and (\ref{eq2}), one finds that the resulting equation for $\dot z$ is amusingly integrable to give
\be
\sqrt{1+\dot z^2}-\sqrt{f+\dot z^2}=C z^4\,,
\ee
with a constant of motion $C>0$, and hence we obtain
\be
\dot z=\sqrt{\left({Cz^4\over 2}+{m\over 2C}-{Q^2 z^2\over 3C}\right)^2-1}\,,
\ee
which can be solved numerically given the constant $C$ which should be determined from the initial conditions.
Once $z(\tau)$ is found, $t_{U,L}(\tau)$ and $p(z)$ can be found subsequently. $p(z)$ turns out to be especially simple
\be
p(z)={C z^4\over 8\pi G_5 }\,.
\ee
We are interested in expressing the falling trajectory in terms of the boundary time $t_U$ that can be identified
with the time measured in QCD, 
\be
z(t_U)=z\left(\tau(t_U)\right)\,,
\ee
so that we can discuss the thermalization history measured in the QCD time. A short algebra gives us the equation
\be
{dz\over dt_U}=f(z)\sqrt{{\left({Cz^4\over 2}+{m\over 2C}-{Q^2 z^2\over 3C}\right)^2-1\over
\left({Cz^4\over 2}+{m\over 2C}-{Q^2 z^2\over 3C}\right)^2-1+f(z)}}\,,\label{v_shell}
\ee
which can be readily solved numerically.

\begin{table}[t]
\begin{center}
\begin{tabular}{c|c|c|c|c}
\hline $\mu_A$ (MeV)& $z_H$ (fm) &$m$ (${\rm fm}^{-4}$) & $Q$ (${\rm fm}^{-3}$) & $C$ (${\rm fm}^{-4}$)\\\hline
50&0.209&526.8&5.82&264.3\\
100&0.208&535.7&11.7&268.6\\
200&0.206&571.8&23.9&286.0\\\hline
\end{tabular} 
\caption{The parameters of the numerical solutions for {\bf RHIC} with the late-time temperature $T=300$ MeV and several exemplar values of $\mu_A$. }\label{tab1}
\end{center}
 \end{table} \begin{table}[t]
\begin{center}
\begin{tabular}{c|c|c|c|c}
\hline $\mu_A$ (MeV)& $z_H$ (fm) &$m$ (${\rm fm}^{-4}$) & $Q$ (${\rm fm}^{-3}$) & $C$ (${\rm fm}^{-4}$)\\\hline
50&0.157&1660.9&10.3&832.7\\
100&0.156&1676.7&20.7&840.4\\
200&0.155&1740.3&42.0&871.2\\\hline
\end{tabular} 
\caption{The parameters of the numerical solutions for {\bf LHC} with the late-time temperature $T=400$ MeV and several exemplar values of $\mu_A$. }\label{tab2}
\end{center}
 \end{table} 
Let us discuss the initial conditions in our numerical solutions that are meaningful in heavy-ion experiments at {\bf RHIC} and {\bf LHC}.
One can conveniently measure the time and space distances in terms of fm (Fermi), and the energy in terms of
${\rm fm}^{-1}=197$ MeV. The relation $z_H={1\over \pi T}$ in the neutral blackhole solution ($Q=0$) comes from the Euclidean geometry stating that $z_H={1\over\pi}\beta$ where $\beta$ is the period of the compactified Euclidean time. Since this period (the inverse temperature) is now measured in units of fm, one can also measure the holographic coordinate $z$ in fm. According to the holographic principle, $z$ maps to the inverse energy scale in the QCD which is also measured in fm, but what is not fixed a priori is a possible numerical rescaling between $z$ measured in fm and the inverse energy scale in QCD also measured in fm.
Guided by the relation $z_H={1\over\pi T}$ for the neutral blackhole ($Q=0$), we will assume the relation between $z$ and the QCD energy scale $E$ as\footnote{This relation in fact depends on what probe we are looking at in the holography. For example, for fundamental quark flavor, the relation between the quark mass and the position $z$ of the probe brane contains an extra factor $\sqrt{g_{YM}^2 N_c}$. Since the blackhole describes deconfined degrees of freedom of gluons, and we are mainly interested in thermalization of gluonic degrees of freedom in our description, the mapping (\ref{map}) guided by the blackhole seems appropriate for our purpose.}
\be
z={1\over\pi E}\,,\label{map}
\ee
with both sides being measured in fm.
\begin{figure}[t]
	\centering
	\includegraphics[width=7cm]{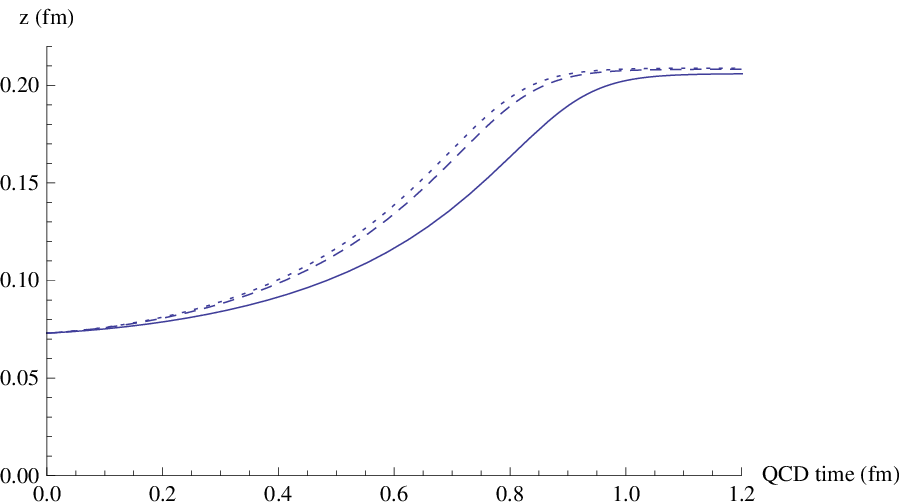}\includegraphics[width=7cm]{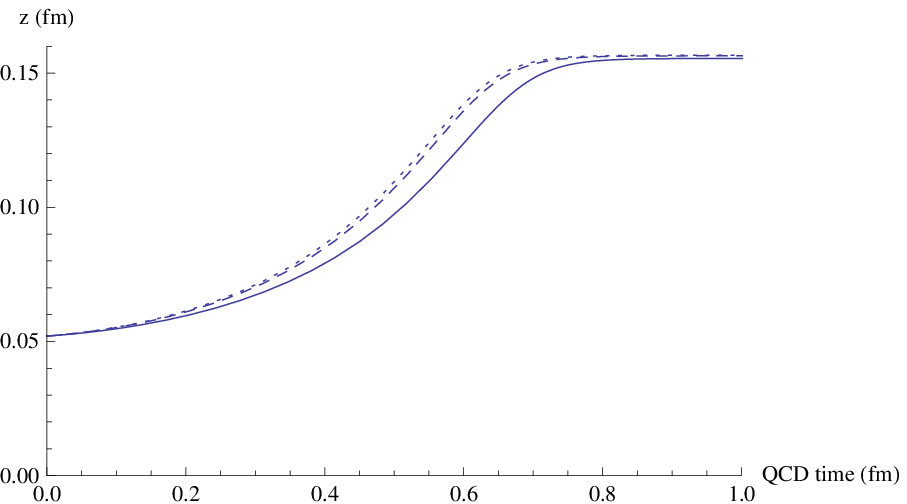}
		\caption{ The thermalization history of the falling mass shell for {\bf RHIC} (left) and {\bf LHC} (right). 
		The late-time temperature is $T=300\, (400)$ MeV for {\bf RHIC} ({\bf LHC}), and the axial chemical potentials are $\mu_A=50$ MeV (dotted), $\mu_A=100$ MeV (dashed), and $\mu_A=200$ MeV (solid). We observe that the system thermalizes mostly by $t\lesssim 1$ fm, and the (axial) charge delays the thermalization.
	\label{fig1}}
\end{figure}
The natural initial condition for the out-of-equilibrium plasma created right after the collision of two heavy-ions
is characterized by the saturation scale $Q_s$, which governs the initial gluon distributions. Roughly speaking, gluons with momenta less than $Q_s$ are densely saturated in the distribution, whereas the states with higher momenta than $Q_s$ are under-occupied, so that $Q_s$ sets a nice boundary between the different UV and IR behaviors. Therefore, we naturally set our initial condition of the falling mass shell to be
\be
z\left(t_U=0\right)=z_i={1\over \pi Q_s}\,,\quad \dot z\left(t_U=0\right)=0\,.
\ee
For {\bf RHIC}, we take $Q_s=0.87$ GeV=4.42 ${\rm fm}^{-1}$, and for {\bf LHC} we have $Q_s=1.23$ GeV=6.24 ${\rm fm}^{-1}$.
To fix $m$ and $Q$ in the solutions, we use the late time temperature $T=300$ MeV for {\bf RHIC} and $T=400$ MeV for {\bf LHC} and several exemplar values for $\mu_A$ using the relations between them and $(m,Q)$ given by (\ref{Tmu}). These data and the above initial conditions are enough
to determine the integration constant $C$ and the unique numerical solution.
See Table \ref{tab1} and Table \ref{tab2}. 

In Figure \ref{fig1} we show the time history of falling mass shell trajectory in QCD time $t_U$ for a few exemplar values of $\mu_A=50,100,200$ MeV. By the time $t\lesssim 1$ fm, the system thermalizes mostly, and we observe that the finite (axial) charge density somewhat delays the thermalization. From gravity point of view, this can be understood as Coulomb repulsion of axial charge acting against the gravitational attraction in the formation of charged blackhole.

\section{ Global geometry of the solution and the quasi-static approximation \label{penrose}}

\begin{figure}[t]
	\centering
	\includegraphics[width=7.9cm]{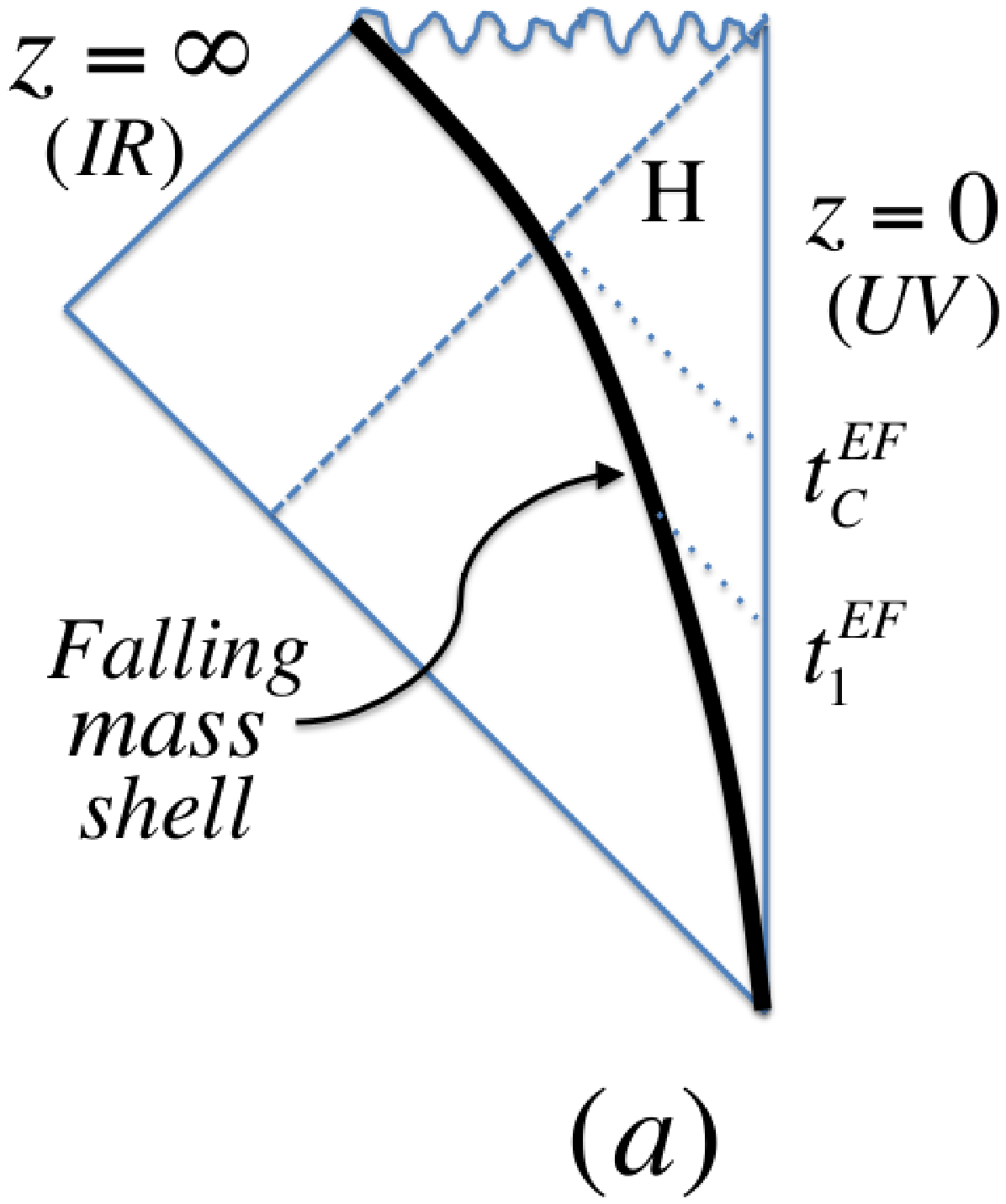}		
	\includegraphics[width=7.9cm]{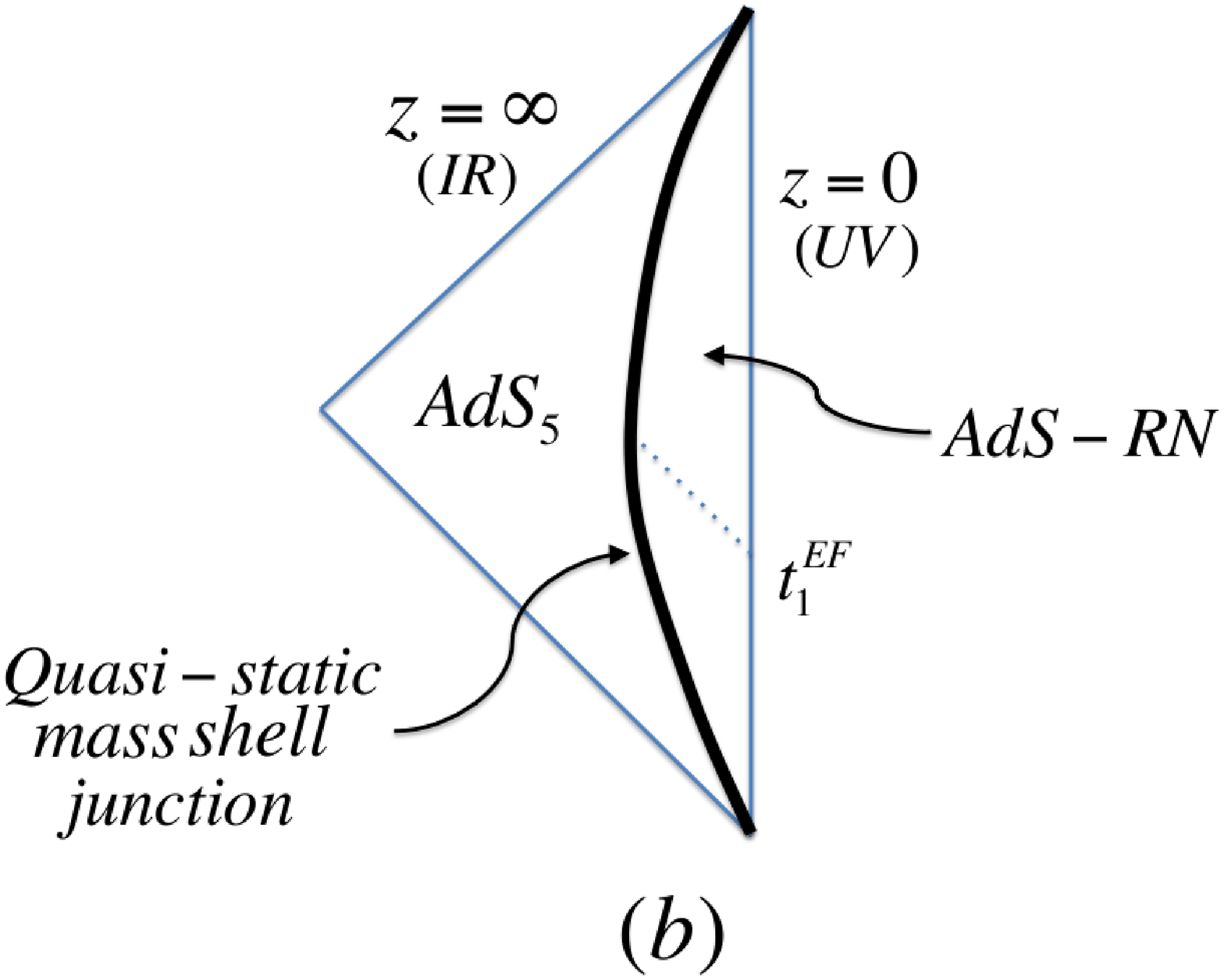}	
	\caption{ (a) The Penrose diagram of the gravitationally collapsing mass shell geometry, (b) The Penrose
diagram of the geometry in the quasi-static approximation.	\label{penrosefig}}
\end{figure}
Before going into the detailed computations of chiral magnetic conductivity and chiral magnetic wave in the solutions obtained in the previous section, it is useful to understand the global structure of the geometry of the solutions and the quasi-static approximation we are going to use.
This will help us to understand the applicability and the limitation of the quasi-static approximation: the quasi-static approximation will be fine far away from equilibrium, but will not be trustable when the mass shell is close enough to the equilibrium horizon. 

The Penrose diagram of the falling mass shell solution in the previous section is given in Figure \ref{penrosefig}(a). The mass shell (the black thick line) falls into a singularity at $z=\infty$, and it
crosses an event horizon (denoted as H) in a finite Eddington-Finkelstein time $t^{EF}_c$.
Note that the Eddington-Finkelstein time is better suited to correctly capture the causal 
structure in the geometry: a light signal sent from the UV boundary $z=0$ propagates into the bulk geometry
whose trajectory is a line of constant Eddington-Finkelstein time by definition (the dashed line with $t_1^{EF}$). Since any response should remain inside a causal light-cone defined by these light geodesics,
these constant Eddington-Finkelstein time lines set a causal structure of the response functions.
For example, it is clear that any signal that is sent after $t^{EF}_c$ (the time when the mass shell crosses the event horizon) would feel the full presence of the event horizon, so that the system after $t^{EF}_c$ will be a fully thermalized plasma, that is, any response functions after $t^{EF}_c$ will precisely be equal to the thermal response functions determined by the event horizon.
The signals sent before $t_c^{EF}$ may see the presence of the falling mass shell above the horizon (like the one with $t_1^{EF}$), so the responses from those signals can include non-equilibrium features.
This means that $t^{EF}_c$ can be interpreted as the thermalization time of the falling mass shell solution. 

The relation between the time $t_U$ in the previous section (QCD time) and the Eddington-Finkelstein time $t^{EF}$ is easily found as
\be
t^{EF}=t_U-\int^z_0 dz'\,{1\over f(z')}\,,
\ee
so that they agree at the UV boundary $z=0$. The AdS-RN metric above the shell looks in terms of $t^{EF}$ as
\be
ds^2={1\over z^2}\left(-f(z) (dt^{EF})^2-2 dt^{EF} dz+(d\vec x)^2\right)\,.
\ee
The falling trajectory in the previous section is given in terms of $t^{EF}$ as
\be
{dz\over dt^{EF}}={f(z)\sqrt{\left({Cz^4\over 2}+{m\over 2C}-{Q^2 z^2\over 3C}\right)^2-1}\over
\sqrt{\left({Cz^4\over 2}+{m\over 2C}-{Q^2 z^2\over 3C}\right)^2-1+f(z)}-\sqrt{\left({Cz^4\over 2}+{m\over 2C}-{Q^2 z^2\over 3C}\right)^2-1}}\,.
\ee
Note that in terms of the original time $t$, it takes an infinite time for the mass shell to cross the event horizon at $z=z_H$, but it is a coordinate artifact. A finite $t^{EF}_c$ is manifested in a less obvious way in the $t$ coordinate: it is the critical time after which the signal of light cannot catch the falling shell \cite{Erdmenger:2012xu}. We stress, however, that the above argument does not take into account spatial extension of the background and the probe. Space-like probes such as strings and Wilson lines can easily give a thermalization time larger than $t^{EF}_c$, see for example \cite{Balasubramanian:2010ce}.

Figure \ref{penrosefig}(b) shows the Penrose diagram of the quasi-static approximation geometry:
the static mass shell (the thick black line) sitting at a constant radius $z=z_s$ borders the AdS-RN geometry
above the shell ($z<z_s$) and the pure AdS below the shell ($z>z_s$). 
We see that for a fixed time $t_1^{EF}$, the difference between the full space (Figure \ref{penrosefig}(a)) and the quasi-static geometry may be small around the region of the constant Eddington-Finkelstein time $t_1^{EF}$ geodesic, if we can safely neglect the velocity of the falling of the mass shell at that moment.
This is the case where the quasi-static approximation is applicable, and this happens when the mass shell is far away from the horizon describing far out-of-equilibrium situations.
However, as the time becomes closer to the critical time $t^{EF}_c$, it is clear that the quasi-static geometry cannot capture the process of thermalization: there is no counterpart of the true event horizon in Figure \ref{penrosefig}(b). The IR horizon at $z=\infty$ is non-thermal. The quasi-static geometry at $z_s=z_H$
is a singular ill-defined geometry where the blackhole horizon and the IR horizon overlap with a zero proper length separation.

\section{Out-of-equilibrium chiral magnetic conductivity}

Given the time dependent backgrounds obtained in the previous section as the holographic description of out-of-equilibrium plasma with a finite axial charge density, it is interesting to see how the properties of the plasma evolve in time. We are interested in the charge transports originating from triangle anomaly in the presence of the magnetic field, and in this section we will treat the magnetic field as a probe to the axially charged plasma, and compute the corresponding Chiral Magnetic current, generalizing the results of \cite{Yee:2009vw} to out-of-equilibrium case. Although the most precise way of studying the problem would be to solve the time-dependent partial differential equations of the system, we will simplify the problem by approximating the falling mass shell to be quasi-static compared to the time-scales of the probes, so that we can solve the time-independent ordinary differential equations instead. We will discuss the regime of validity of the quasi-static approximation in our results later.

Treating the magnetic field as a probe, we will compute the chiral magnetic conductivity, $\sigma_\chi$, defined by
\be
\vec J_{EM}=\sigma_\chi(\omega) \vec B(\omega)\,,
\ee
where the magnetic field (probe) has a definite frequency $\omega$. One naturally expects that our results for very low $\omega$ would not be consistent with the quasi-static approximation, and we will specify precisely where we can trust the results shortly. As the position of the quasi-static mass shell changes in time, the chiral magnetic conductivity also evolves in time. Combining with the time history of the falling mass shell in the previous section then allows us to discuss the time-evolution of the chiral magnetic conductivity in realistic conditions relevant for {\bf RHIC} and {\bf LHC}.

We turn on a time-dependent magnetic field of frequency $\omega$ as a linearized probe, and try to find the response of the system
given by the (quasi-static) mass shell geometry with an axial charge density,
\bear
ds^2_{U}&=&{dz^2\over f(z) z^2}-{f(z)\over z^2}dt^2 + {\left(d\vec x\right)^2\over z^2}\quad({\rm upper\,\, region}: z<z_s)\,,\nonumber\\
ds^2_{L}&=&{dz^2\over  z^2}-{dt_L^2 \over z^2}+ {\left(d\vec x\right)^2\over z^2}\quad({\rm lower\,\, region}: z>z_s)\,,
\eear
where $z=z_s$ is the (quasi-static) location of the mass shell. Note that we have used the notation $t$ for the upper part of the metric since it is identified with the QCD time. The two times $t$ and $t_L$ are matched at $z=z_s$ by
\be
\sqrt{f(z_s)}t=t_L\,,\label{tandtb}
\ee 
in order for the whole metric to be continuous, which is the quasi-static limit of the Israel junction condition.

The matching relation (\ref{tandtb}) in frequency space becomes
\be
\omega=\sqrt{f(z_s)}\omega_L\,,
\ee
which will be used in solving the equations in the frequency space.
 Inspecting the linearized equations of motion
from (\ref{mastereom}), one can easily find that the equation for the vector gauge field $A_V$ decouples from those of the metric and the axial gauge field $A_A$, and since the current and the magnetic field of our interests
are all vector quantities, it is enough to consider that equation only,
\be
\partial_N\left(\sqrt{-g_5}\left(F_V\right)^{MN}\right)-{3\kappa\over 2}\epsilon^{MNPQR}\left(F_A^{(0)}\right)_{NP}\left(F_V\right)_{QR}=0\,,\label{maineq}
\ee
where the vector fields appearing represent linearized fluctuations from our background solution in the previous section, and $F_A^{(0)}$ is the background value of the axial gauge field in the solution, given by
\be
F_A^{(0)}=dA_A^{(0)}=-2zQ dz\wedge dt\quad({\rm upper\,\, region})\,,\quad F_A^{(0)}=0\quad({\rm lower\,\, region})\,.
\ee
Noting that the shell is vector charge neutral, the natural junction condition for the gauge field is the continuity of its value and normal derivative. We choose to work in the gauge $A_z=0$ for both upper and lower regions. From the continuity of the value and the normal derivative, we require $[A_Mdx^M]=[F_{MN}n^Mdx^N]=0$ where $n^M$ is the unit normal vector to the shell. We substitute $dx^M=\frac{\del x^M}{\del\xi^i}d\xi^i$ and noting that $d\xi^i$ can be arbitrary, we end up with
\be
[A_M\frac{\del x^M}{\del\xi^i}]=[F_{MN}n^M\frac{\del x^N}{\del\xi^i}].
\ee
In the quasi-static approximation, we simply set $\dot{z}=0$ in (\ref{nul}), and from the above
junction conditions, we find
\be
A^U_t=\sqrt{f(z_s)} A^L_{t_L}\,,\quad A^U_i=A^L_i\,\,(i=1,2,3)\,,\label{june1}
\ee
whereas the continuity of the normal derivatives gives us
\be
\partial_z A^U_t=\partial_z A^L_{t_L}\,,\quad \sqrt{f(z_s)}\partial_z A^U_i=\partial_z A_i^L\,.\label{june2}
\ee
We have omitted the subscript $V$ without confusion.
We solve (\ref{maineq}) with the above junction conditions at $z=z_s$.

To introduce a magnetic field along, say, $x^3$ direction, we consider a fluctuation of $A_2$ with a momentum along $x^1$ to have a non-zero $F_{12}=B_3$,
\be A_2(t,\vec x,z)=A_2(z) e^{-i\omega t+ikx^1}\,,
\ee
and the consistency of the equation of motion (\ref{maineq}) necessitates the introduction of $A_3$ fluctuation as well,
\be
A_3(t,\vec x,z)=A_3(z) e^{-i\omega t+ikx^1}\,.
\ee
This coupling between $A_2$ and $A_3$ is via the Chern-Simons term, and indeed we will obtain the non-zero chiral magnetic current along $x^3$ (the direction of the magnetic field) from the induced $A_3$ fluctuation.
Other components of the gauge field can be turned off consistently. The equations of motion are explicitly given as
(note our convention $\epsilon^{zt123}=1$), 
\bear
\partial_z\left({f\over z}\partial_z A_2^U\right)+{1\over z}\left({\omega^2\over f}-k^2\right)A_2^U
+12 i\kappa Q z k A_3^U&=&0\,,\nonumber\\
\partial_z\left({f\over z}\partial_z A_3^U\right)+{1\over z}\left({\omega^2\over f}-k^2\right)A_3^U
-12 i\kappa Q z k A_2^U&=&0\,,\nonumber\\
z\partial_z^2 A_{2}^L-\partial_z A_{2}^L+z\left(\omega_L^2-k^2\right)A_{2}^L&=&0\,,\nonumber\\
z\partial_z^2 A_{3}^L-\partial_z A_{3}^L+z\left(\omega_L^2-k^2\right)A_{3}^L&=&0\,,\label{eomfre}
\eear
where the first two equations are in the upper region and the last two in the lower region. In the lower region, the equations are easily solved by Hankel functions, and we require the infalling boundary condition for the physical retarded response functions. In the upper region, one has to solve the equation numerically.
Since we would like to turn on the external magnetic field along $x^3$ direction, the $A_2^U$ field should have a near boundary expansion close to $z=0$ as
\be
A^U_2(z)=A^{(0)}_2+A^{(2)}_2 z^2+A_2^h z^2\log z+\cdots\,,
\ee
and the external magnetic field is identified as
\be 
eB=F_{12}^{(0)}=ik A_2^{(0)}\,.
\ee
The $A_3^U$ field should not have any boundary value by the choice of the boundary condition, so its near boundary expansion should be
\be
A^U_3(z)=A^{(2)}_3 z^2+A_3^h z^2\log z+\cdots\,.
\ee
The infalling boundary condition in the lower region and the above near $z=0$ boundary condition in the upper region uniquely determine the full solution, which is linear in the value $A_2^{(0)}$ (and hence the magnetic field) that sets the overall normalization. Note that we have to match the solutions in the two regions via the junction conditions (\ref{june1}) and (\ref{june2}). Once the solution is found given the normalization set by $A_2^{(0)}$,
the current along $x^3$ direction which is our chiral magnetic current along the direction of the magnetic field is obtained as
\be
J^3_{EM}=eJ_V^3={e\over 4\pi G_5}A^{(2)}_3\,,\label{j3}
\ee
so that the chiral magnetic conductivity is given by
\be
\sigma_\chi={J^3\over B}={e^2\over 4 \pi G_5} {A_3^{(2)}\over ik A_2^{(0)}}\,,\label{chimc}
\ee
which is well-defined independent of the normalization of the solution. 

The prescription (\ref{j3}) needs some explanations. 
In the careful holographic renormalization of Einstein-Maxwell-Chern-Simons theory \cite{Sahoo:2010sp}, the near boundary expansion of the gauge field is given by $A_\mu=A_\mu^{(0)}+ A_\mu^{(2)}z^2+A^h_\mu z^2\log z^2+\cdots$. Note that our action density (\ref{ad}) has already taken into Bardeen counter term. The current expectation value can be obtained from functional derivative of the action as:
\be
J_\mu={1\over 4\pi G_5}\left(A_\mu^{(2)}+A^h_\mu\right)+{3\kappa\over 8\pi G_5}\epsilon^{\mu\nu\alpha\beta}\left((A_A^{(0)})_\nu(F_V^{(0)})_{\alpha\beta}\right)\,.
\ee
The last contribution from the Chern-Simons term needs a special care. To obtain physical chiral magnetic effect, one needs to distinguish axial chemical potential $\mu_A$ and boundary value of axial gauge field $A_A^{(0)}$ \cite{Gynther:2010ed,Kharzeev:2011rw}. In Minkowski signature black-hole solution, the time component of $A_A$ does not need to vanish at the horizon without causing any singularity problem \cite{Kharzeev:2011rw}. The boundary value of the axial gauge field $A_A^{(0)}$ is clearly zero in real physical configuration created in heavy-ion collisions, while the chemical potential is simply defined as a work needed to bring a unit charge from infinity to the plasma, so that the two things are different. For this reason we have chosen the gauge field configuration $A_A$ to have vanishing boundary value, but correspond to a finite chemical potential.
Having zero $A_A^{(0)}$ in our plasma in heavy-ion collisions gives no additional contribution to (\ref{j3}) from the Chern-Simons term, and 
it does not affect our formula (\ref{j3}) for the $J^3$.

We are interested in the homogeneous magnetic field with a finite frequency, so we would like to consider $k\to 0$ limit while $B$ is fixed. This limit can be achieved in the following way \cite{Yee:2009vw}.
Looking at the equations of motion (\ref{eomfre}), the terms originating from the Chern-Simons term that mix $A_2$ and $A_3$ are linear in $k$, so that one naturally expects that the induced $A_3$ fluctuation from the source of $A_2$ (the magnetic field)
will be linear in $k$ in $k\to 0$ limit. Therefore, the chiral magnetic conductivity from (\ref{chimc}) has a well-defined finite value in $k\to 0$ limit. Since we are only interested in the linear $k$ dependence in $A_3$,
the other $k^2$ terms in (\ref{eomfre}) are not relevant, and can be neglected. These considerations lead to
expanding the solution in powers of $k$ as
\be
A_2(z)=a_2(z)+{\cal O}(k^2)\,\quad A_3(z)=k a_3(z)+{\cal O}(k^3)\,,
\ee
where $a_2$ and $a_3$ satisfy the equations
\bear
\partial_z\left({f\over z}\partial_z a_2^U\right)+{1\over z}{\omega^2\over f}a_2^U
&=&0\,,\nonumber\\
\partial_z\left({f\over z}\partial_z a_3^U\right)+{1\over z}{\omega^2\over f}a_3^U
-12 i\kappa Q z  a_2^U&=&0\,,\nonumber\\
z\partial_z^2 a_{2}^L-\partial_z a_{2}^L+z\omega_L^2a_{2}^L&=&0\,,\nonumber\\
z\partial_z^2 a_{3}^L-\partial_z a_{3}^L+z\omega_L^2a_{3}^L&=&0\,,\label{eomfre2}
\eear
with the same junction conditions (\ref{june1}) and (\ref{june2}).
The frequency dependent chiral magnetic conductivity then becomes
\be\label{chimc_k}
\sigma_\chi(\omega)=-i {e^2\over 4\pi G_5}{a_3^{(2)}\over a_2^{(0)}}\,,
\ee
with a similar near boundary expansion as before,
\be
a_2^U=a_2^{(0)}+a_2^{(2)}z^2+\cdots\,,\quad a_3^U=a_3^{(2)}z^2+\cdots\,.
\ee
The use of (\ref{chimc}) requires that $A_3^U$ tends to zero as it approaches the boundary. However, fine tuning the boundary value is not numerically convenient.
In appendix A, we show how to calculate numerically both chiral magnetic conductivity and electric conductivity from the solutions with non-vanishing boundary value of $A_3^U$.

\begin{figure}[t]
	\centering
	\includegraphics[width=7cm]{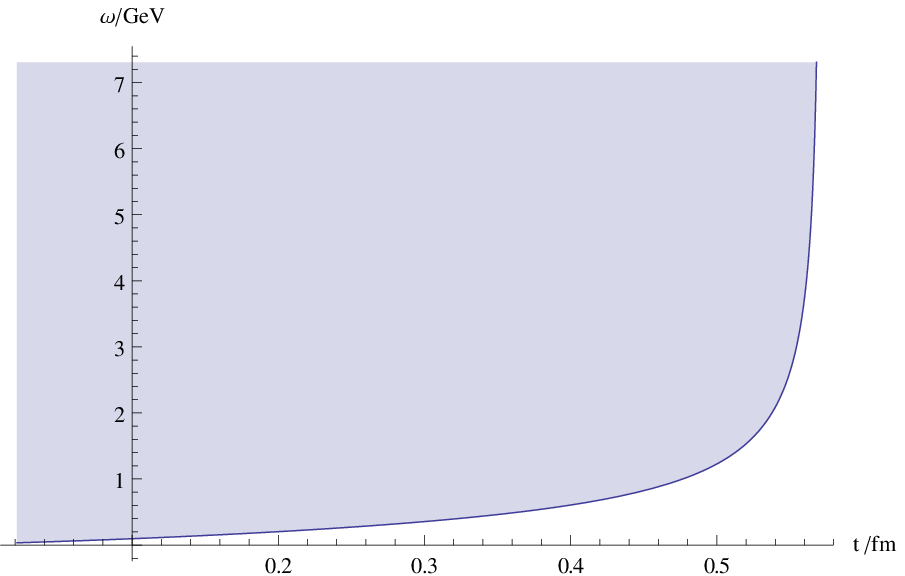}\includegraphics[width=7cm]{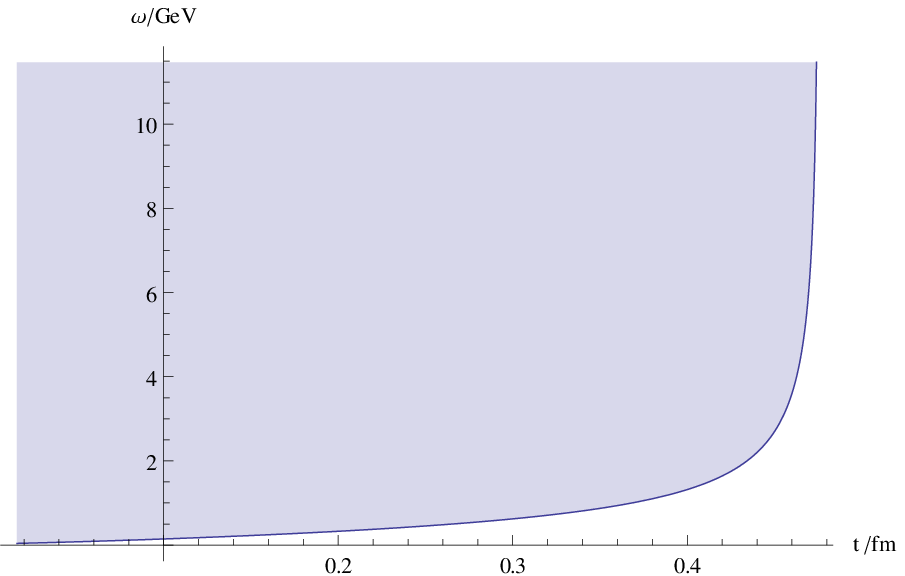}
		\caption{ The region of validity of quasi-static approximation in the frequency space as a function of time for RHIC ($T=300$ MeV, $\mu_A=50$ MeV) and LHC ($T=400$ MeV, $\mu_A=50$ MeV). The shaded region above the curve is consistent with the quasi-static approximation.
	\label{reg_valid}}
\end{figure}

As remarked previously, the quasi-static approximation has its limitation. It is valid when the speed of the probe, in this case speed of light for the gauge field, is much greater than the falling speed of the shell. As the shell approaches the ``horizon'', both the shell and the speed of light are infinitely red-shifted. We expect the quasi-static approximation to break down as $z\to z_H$ as discussed in section \ref{penrose}. Furthermore, this picture relies on the assumption that we can treat the gauge field as a massless particle. It is justified when the wave length of the gauge field is much shorter than curvature of AdS space. This is given by $\omg z\gtrsim 1$. These provide sufficient conditions for quasi-static approximation. In appendix B, we work out more precise conditions to find that a wide region in the frequency $\omega$ space appears to be consistent with the quasi-static approximation. We simply quote the results here:
\bear
\dot{z}&\ll& \frac{H_0^{(1)}(\omg z/\sqrt{f})}{H_1^{(1)}(\omg z/\sqrt{f})}\sqrt{f+\dot{z}^2}, \nonumber\\
\sqrt{f+\dot{z}^2}&\gg& \frac{H_0^{(1)}(\omg z/\sqrt{f})}{H_1^{(1)}(\omg z/\sqrt{f})}\dot{z}.
\eear
Figure \ref{reg_valid} shows the region of validity for the quasi-static approximation. Generically, the quasi-static approximation corresponds to probing the evolving medium with a plane wave, which has infinite resolution $\Dlt\omg=0$ in frequency, but vanishing resolution $\Dlt t=\infty$ in time. We know by uncertainty principle $\Dlt\omg\Dlt t\ge1/2$. For a medium evolving sufficiently slow in time, $\Dlt t$ can be made very large, which allows for a small $\Dlt\omg$. This is the way how the quasi-static approximation works. The breaking down of quasi-static approximation at late time seems to suggest that the evolution of medium becomes faster at late stage of thermalization, while a naive expectation from slow motion of the shell near horizon that would lead to the opposite conclusion is illusionary, as it is also clear in the Penrose diagrams in section \ref{penrose}.

With the falling trajectory obtained in the previous section, $z_s=z(t)$, one can discuss how $\sigma_\chi(\omega)$ changes in QCD time $t$ in our quasi-static approximation. 
Note that the chiral magnetic conductivity $\sigma(\omega)$ in general has both real and imaginary parts, and we would like to parametrize it by the magnitude $|\sigma_\chi(\omega)|$ and the response time delay $\Delta t(\omega)=arg(\sig_\chi(\omg))/\omg$, defined by
\be
J^3e^{-i\omg t}=\sig_\chi(\omg) Be^{-i\omg t}=|\sig_\chi(\omg)| B^{-i\omg(t+\Delta t(\omega))}.
\ee
\begin{figure}[t]
	\centering
	\includegraphics[width=7cm]{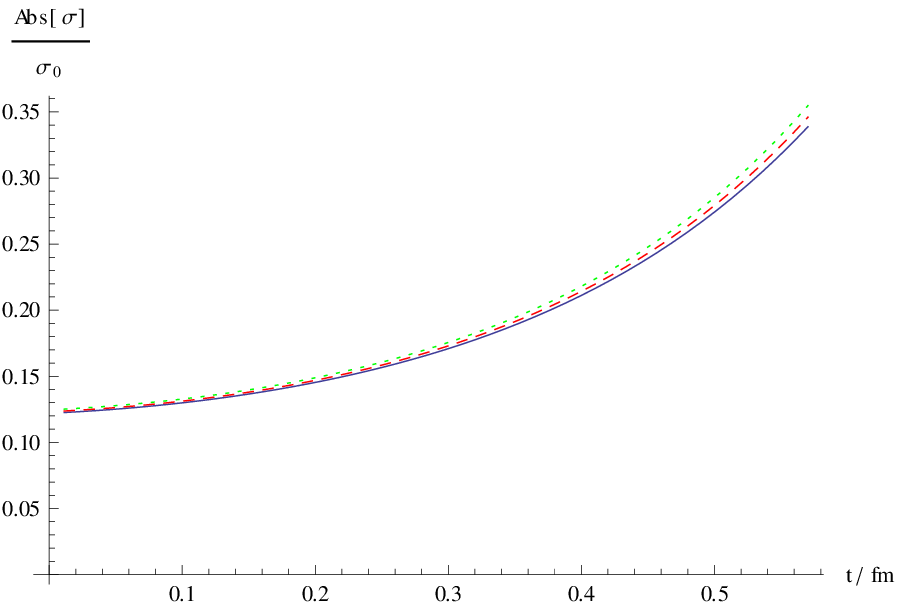}\includegraphics[width=7cm]{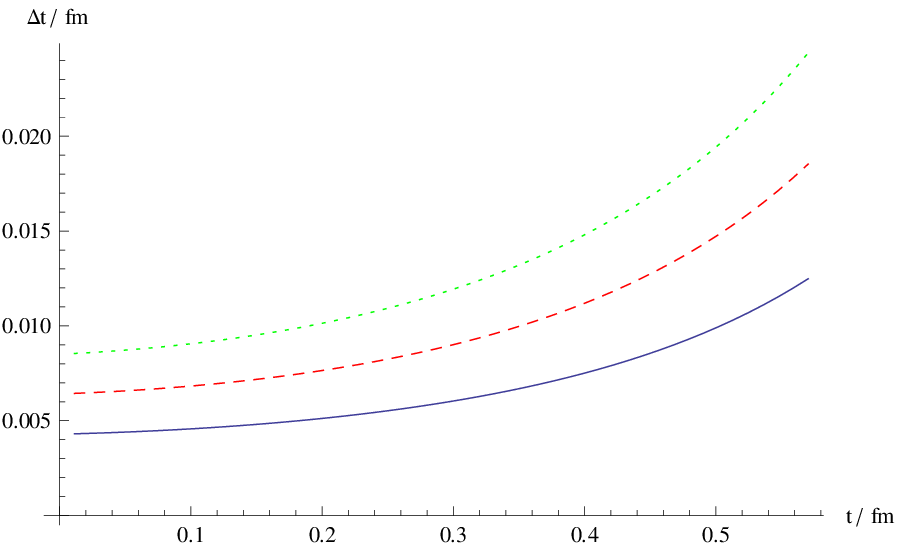}
		\caption{ The chiral magnetic conductivity as a function of thermalization history for different frequencies: $\omg=200$ MeV(blue solid), $\omg=300$ MeV(red dashed) and $\omg=400$ MeV(green dotted).  The thermalization history of the falling mass shell is for {\bf RHIC} with a final temperature $T=300$ MeV and $\mu_A=50$ MeV. The left plot shows the evolution of the magnitude of chiral magnetic conductivity and the right plot shows the time delay of the response. 
	\label{cmc_omg}}
\end{figure}
As an example, we plot in Figure \ref{cmc_omg} the time evolution of chiral magnetic conductivity characterized by the magnitude and the response time delay for three particular values of $\omega$ with a fixed $\mu_A$.  We plot the same quantities in Figure \ref{cmc_mu} for three values of $\mu_A$ with a fixed value of $\omega$. We present the results with respect to the equilibrium zero frequency value of chiral magnetic conductivity
\be
\sigma_0\equiv -{3\kappa e^2\over 4\pi G_5}\mu_A={e^2 N_c\over 2\pi^2}\mu_A\,,
\ee
where the last equality comes from the relation (\ref{kappa}): $\kappa=-{2 G_5 N_c\over 3\pi}$.
Several conclusions can be drawn from our results: 
\begin{figure}[t]
	\centering
	\includegraphics[width=7cm]{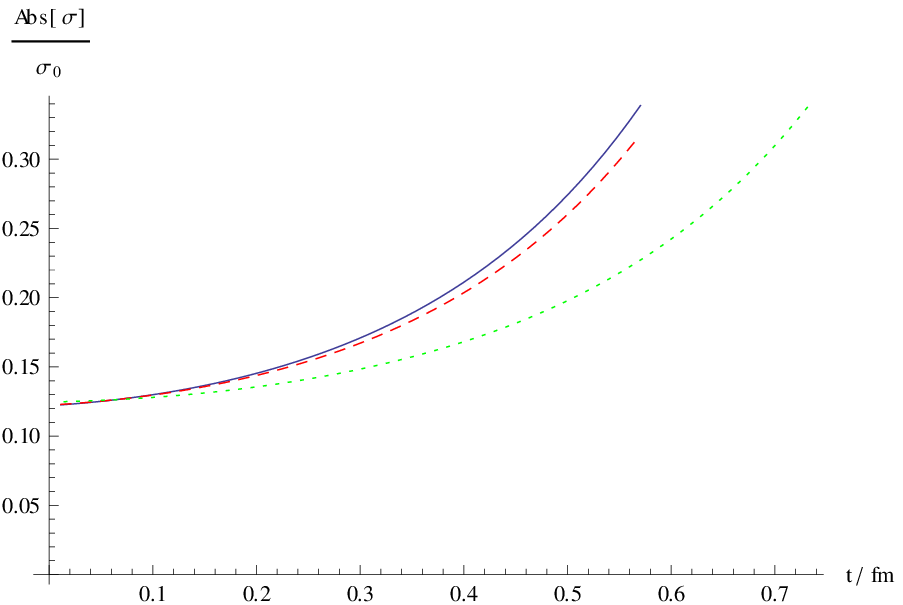}\includegraphics[width=7cm]{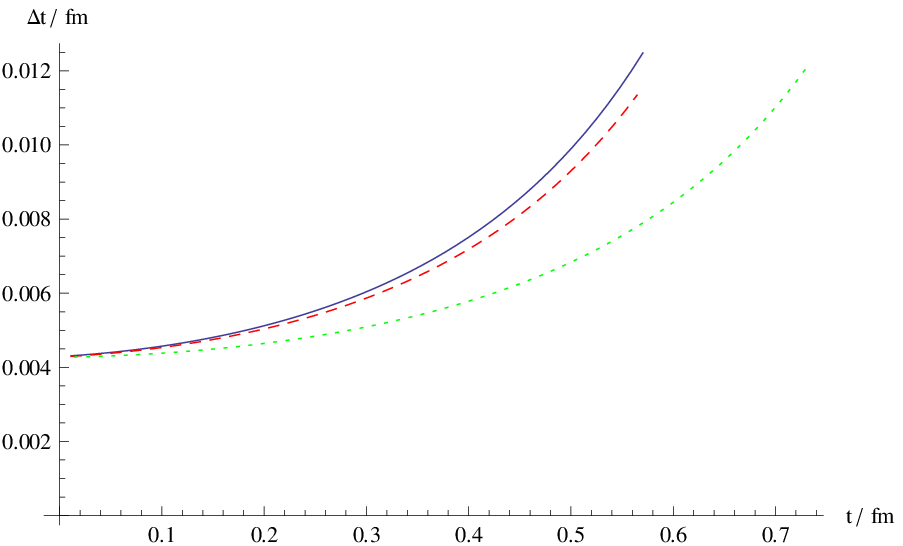}
		\caption{ The chiral magnetic conductivity as a function of thermalization history at a fixed frequency $\omg=200$ MeV for different axial chemical potentials: $\mu_A=50$ MeV(blue solid), $\mu_A=100$ MeV(red dashed) and $\mu_A=200$ MeV(green dotted). The thermalization history of the falling mass shell is for {\bf RHIC} with a final temperature $T=300$ MeV. The left plot shows the evolution of the magnitude of chiral magnetic conductivity and the right plot shows the time delay of the response. 
	\label{cmc_mu}}
\end{figure}

i) the chiral magnetic conductivity, both its magnitude and the time delay, increases in general as the medium thermalizes. The increase of the magnitude is consistent with the naive expectation that as the medium thermalizes, more and more thermalized constituents can participate in the formation of chiral magnetic current. 

ii) From Figure \ref{cmc_omg} we observe that the magnitude of chiral magnetic conductivity changes very little as we vary the frequency of the probe, while increase of the latter does result in longer delay in the response of the medium. This is in contrast to conventional electric property of materials. Simple Drude model of electric conductivity shows that electric field of higher frequency results in lower magnitude of electric conductivity and shorter delay in response. The difference should not be surprising as the non-dissipative chiral magnetic conductivity is of different nature from the dissipative electric conductivity.

iii) Figure \ref{cmc_mu} shows that a larger chemical potential gives a smaller ratio of the magnitude of the conductivity to $\sigma_0$, and a shorter delay in response. However, we should bear in
mind that a larger chemical potential also delays the thermalization time. Note that
we are comparing the conductivity at the same absolute time, which corresponds to less thermalized medium for larger chemical potential. Therefore the results are consistent with the observation i).

\section{Out-of-equilibrium chiral magnetic wave}

In this section, we study out-of-equilibrium property of charge transports originating from triangle anomaly in a different angle: the chiral magnetic wave. The chiral magnetic wave describes how (chiral) charge fluctuations behave in the presence of an external magnetic field which we assume to be static.
In the equilibrium plasma, the chiral magnetic wave has a dispersion relation of the form \cite{Kharzeev:2010gd}
\be
\omega=\mp v_\chi k-iD_L k^2+\cdots\,,\label{cmwdis}
\ee
with the velocity $v_\chi$ being proportional to the magnetic field
\be
v_\chi={N_c e^2 B\over 4\pi^2 \chi}\,,\quad \chi\equiv {\partial  J^0\over\partial\mu}\,,
\ee
and the sign in the first term (the direction of propagation) depends on the chirality of the fluctuations.
Since the chiral magnetic wave is about linearized charge fluctuations, the background plasma can be neutral and we consider a neutral (out-of-equilibrium) plasma in this section for simplicity. 
We are interested in how the dispersion relation of the chiral magnetic wave changes in time in our out-of-equilibrium conditions represented by falling mass shell geometries in the previous sections. The charge neutral background can be easily found by putting $Q=0$ in the previous solutions. 

The dispersion relation of chiral magnetic wave in neutral plasma in equilibrium can be easily found from poles of retarded current-current correlator in Fourier space. However the same procedure does not carry over straightforwardly out of equilibrium. We know that in equilibrium the poles in the complex $\omg$ plane carry the same information as the full retarded function defined on the real $\omg$ axis. This is because the equilibrium retarded correlator has infinite resolution in $\omg$, allowing for an analytic continuation into the complex plane. For plasma out of equilibrium, the retarded correlator in $\omg$ space is only approximately defined with a finite resolution via Wigner functions, and the analytic continuation may not be well justified. Therefore, we stick to work in the real $\omg$ domain of the retarded Green's function, which is more directly relevant to the real-time behavior of fluctuations. 
As in the previous section, we will work in the quasi-static approximation. The lowest frequency wave-like excitation, that we call out-of-equilibrium chiral magnetic wave, will be identified as a peak in the imaginary part of the correlator (spectral function) below the lightcone $\omg<k$. Higher excitations will in general appear above the lightcone $\omg>k$. We will trace the time evolution of the identified out-of-equilibrium chiral magnetic wave peak.

To compute the spectral function in the falling mass shell geometry, we turn on a static, homogeneous magnetic field along $x^3$ direction $\vec B=B\hat x^3$. We consider a weak magnetic field without backreaction to the shell geometry. A constant magnetic field is a trivial solution of the equations of motion. This constitutes our background solution, from which we consider linearized fluctuations of both axial and vector gauge fields $\delta A_{A,V}$ that describe
chiral charge fluctuations in the QCD side (we omit $\delta$ symbol in the below without much confusion).
The linearized fluctuations of gauge fields in fact decouple from those of the metric in the case of neutral background, and this simplification is one reason why we consider neutral plasma in our study.
The linearized equations for the gauge fields from our main equations (\ref{mastereom}) are diagonalized in the chiral basis defined as
\be
A_L\equiv A_V-A_A\,,\quad A_R\equiv A_V+A_A\,,
\ee which represent chiral charge fluctuations.
Explicitly, their equations read as
\be
\partial_N\left(\sqrt{-g_5}(F_{L,R})^{MN}\right)\pm {3\kappa eB}\epsilon^{M12QR}(F_{L,R})_{QR}=0\,,
\ee
where we have used that the background value of $A_{L,R}$ are given by
\be
(F_{L}^{(0)})_{12}=(F_R^{(0)})_{12}=eB\,.
\ee
Since left- and right-handed fluctuations are simply related by $B\to -B$, let us focus on the right-handed fluctuations only (the lower sign in the above equation) and omit the subscript $R$ in the following. 

The chiral magnetic wave is a longitudinal charge-current fluctuation, so we consider a longitudinal momentum $k$ along $x^3$ (the direction of the magnetic field) and turn on $A_t$ and $A_3$ fluctuations in the gauge $A_z=0$,
\be
A_t=A_t(z)e^{-i\omega t+i k x^3}\,,\quad A_3=A_3(z)e^{-i\omega t+i k x^3}\,,
\ee
where other components of the gauge field can be consistently turned off. The equations of motion then become
\bear
{\omega\over z}\partial_z A_t+{kf\over z}\partial_z A_3+6\kappa eB\left(\omega A_3+k A_t\right)&=&0\,,\nonumber\\
-{k\over zf}\left(\omega A_3+k A_t\right)+6\kappa eB\partial_z A_3+\partial_z\left({1\over z}\partial_z A_t\right)&=&0\,,\nonumber\\
-{\omega\over z f}\left(\omega A_3+k A_t\right)-6\kappa eB\partial_z A_t-\partial_z\left({f\over z}\partial_z A_3\right)&=&0\,,
\eear
for the upper region $z<z_s$ and the equations in the lower region $z>z_s$ is the same with $f=1$.
We have to match the upper and lower solutions by the previous junction conditions (\ref{june1}) and (\ref{june2}).
It is more convenient and intuitive to work with a gauge invariant variable corresponding to the electric field along $x^3$ direction defined as
\be
E\equiv k A_t+\omega A_3\,,
\ee
for which the equation simply becomes
\be\label{eom_probe}
\left(E^U\right)''+\left({\omega^2 f'\over f\left(\omega^2-f k^2\right)}-{1\over z}\right)\left(E^U\right)'
+{1\over f}\left({\omega^2\over f}-k^2-\left(6\kappa eBz\right)^2 +{6\kappa eB\omega k zf'\over \omega^2-fk^2}\right)E^U=0\,,\ee
for the upper region $z<z_s$ where $'\equiv {d\over dz}$, and the equation in the lower region is similar with replacing $f=1$ and $\omg_L=\omg/\sqrt{f(z_s)}$.
The junction condition in terms of $E$ is
\be
E^U=\sqrt{f(z_s)} E^L,\; \left(E^U\right)'=\left(E^L\right)'.
\ee
Guided by the equilibrium chiral magnetic wave,
we expect to find a chiral magnetic wave peak in the positive $\omega$ axis when $k>0$ and $B>0$.

We are ready to solve (\ref{eom_probe}) and its counterpart below the shell, with the junction condition on the shell and in-falling boundary condition for $E^L$ at IR infinity. However we see a subtle problem: due to the $(6\ka eBz)^2$ term, the solution to $E^L$ either diverges or decays exponentially at IR infinity for any frequency momentum. Once we choose the exponentially decaying solution which is naturally real, the full solution will be purely real for any $\omg$ and $k$. This means that the imaginary part of the retarded correlator (spectral function) can only have delta-function peaks corresponding to infinitely stable bound states, without any continuum part of our interest that may feature chiral magnetic wave as the system thermalizes. This unphysical drawback seems to appear as a result of our probe limit, where we neglect the backreaction of the $B$-field to the metric. In our theory the metric at IR infinity will necessarily be changed in the presence of any $B$ no matter how small $B$ is \cite{D'Hoker:2009mm}: the IR geometry should be modified to AdS${}_3\times$R${}^2$. Once this has been taken into account, we checked that the correct geometry does allow the (complex-valued) in-falling IR boundary condition. 

We defer a full treatment including the back reaction of the $B$ field to the future, and use instead the following approximation that still captures the main physics effect of the back reacted geometry: below the shell, we introduce an IR cutoff $z_c$ beyond which we drop the term $(6\ka eBz)^2$ such that the solution can be chosen to be in-falling. This mimics the effect of the AdS${}_3\times$R${}^2$ below the IR cutoff. Above the IR cutoff, we reinstate the term $(6\ka eBz)^2$ and find the full solution up to the UV boundary. The cutoff $z_c$ is naturally chosen to be $z_c=1/\sqrt{B}$ where the back reaction starts to be important \cite{D'Hoker:2009mm}. Our treatment is well justified when
 $B\ll T^2$ and the shell is not too close to the horizon.

\begin{figure}[t]
\centering
\includegraphics[width=7cm]{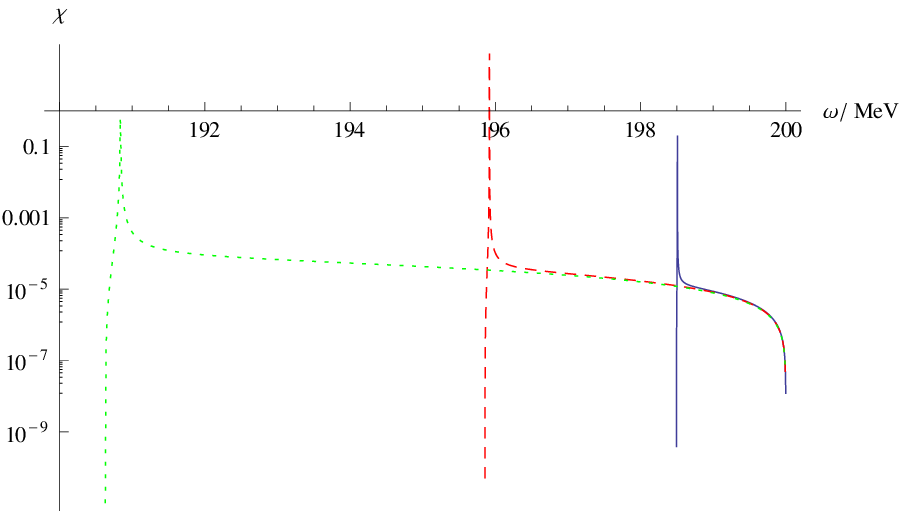}\includegraphics[width=7cm]{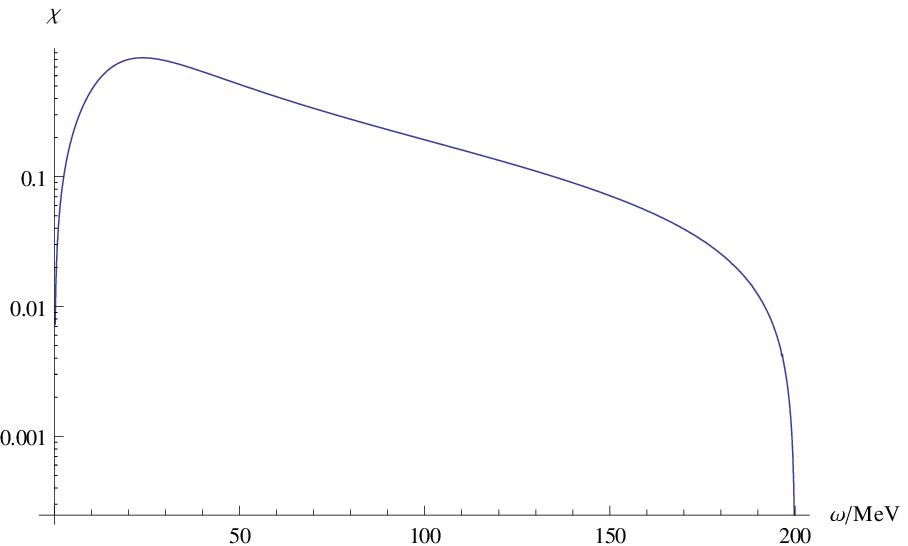}
\caption{The spectral functions in unit of $\frac{(\pi T)^2}{8\pi G_5}$ as a function of $\omg$ at fixed $k=200$ MeV. The magnetic field and temperature are chosen for the RHIC: $B=m_\pi^2$ and $T=300$ MeV, which satisfy the condition $B\ll T^2$ for the justification of our IR cutoff. The left plot shows the snapshots of the spectral functions at $t=0.03$ fm(blue solid), $t=0.38$ fm(red dashed) and $t=0.53$ fm(green dotted). The right plot shows the equilibrium spectral function as a reference.\label{fig_spectral}}
\end{figure}

\begin{figure}[t]
\centering
\includegraphics[width=7cm]{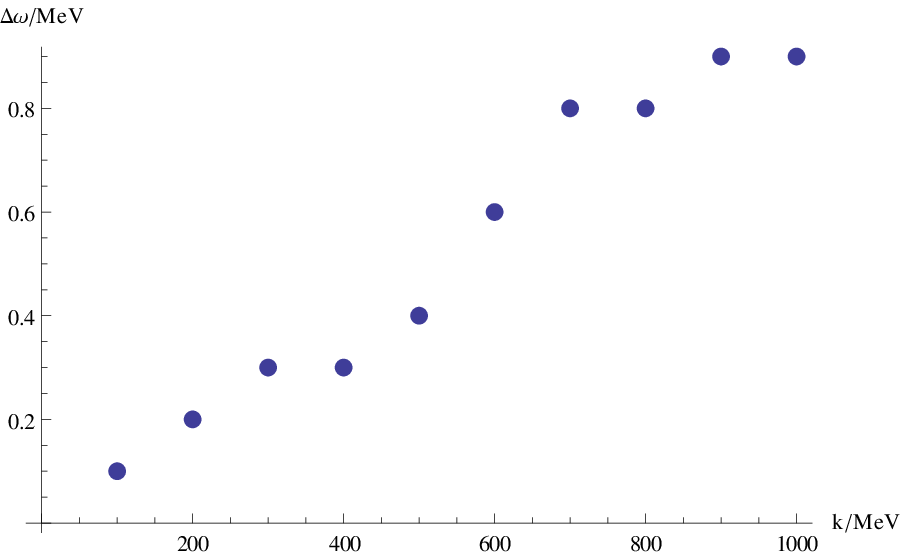}\includegraphics[width=7cm]{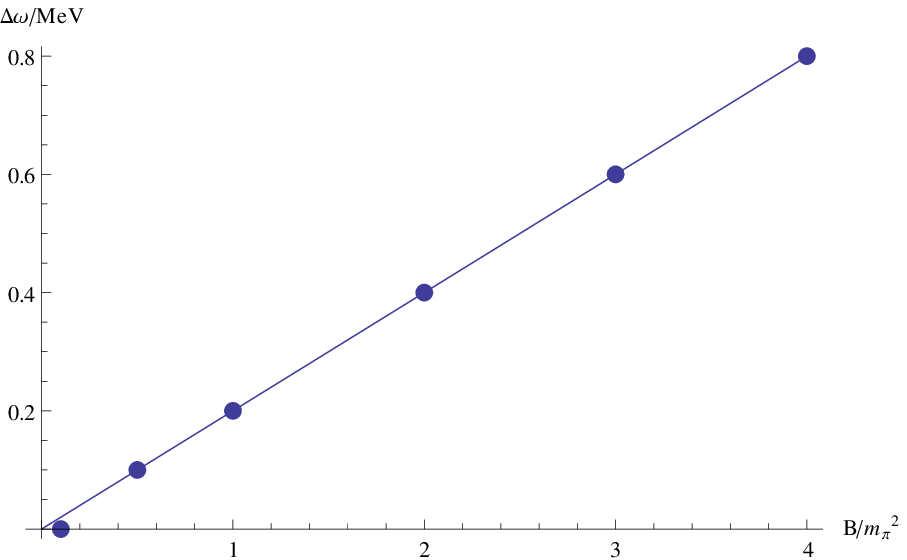}
\caption{Left: $\Dlt\omg$ as a function of $k$ at fixed $B=m_\pi^2$. Right: $\Dlt\omg$ as a function of $B$ at fixed $k=200$ MeV. In both plots, $T=300$ MeV and $t=0.53$ fm.\label{width}}
\end{figure}

\begin{figure}[t]
\centering
\includegraphics[width=7cm]{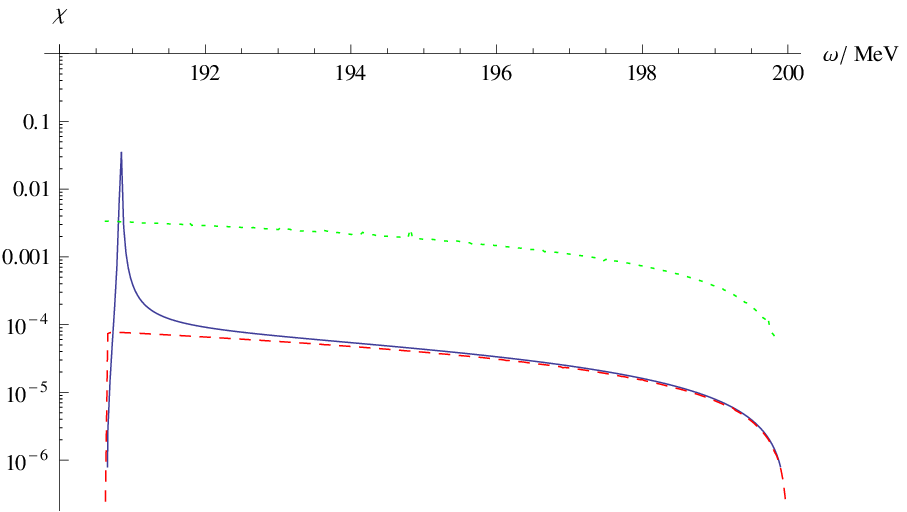}\includegraphics[width=7cm]{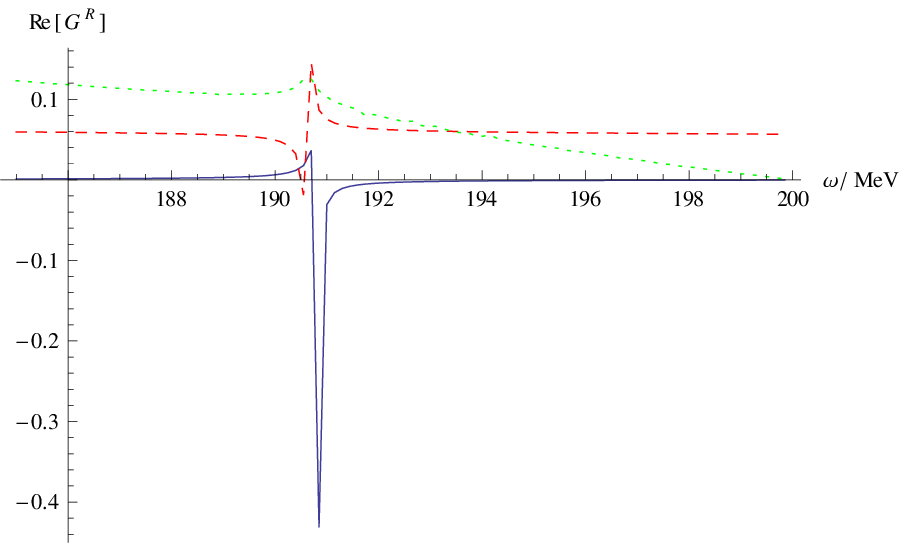}
\caption{Comparison of spectral functions in unit of $\frac{(\pi T)^2}{8\pi G_5}$ taken at $t=0.53$ fm as a function of $\omg$ at fixed $k=200$ MeV and $T=300$ MeV with different magnetic fields: $B=m_\pi^2$ (blue solid), $B=-m_\pi^2$ (red dashed) and $B=0$ (green dotted). The right plot shows the real part of retarded correlator in unit of $\frac{(\pi T)^2}{8\pi G_5}$ with the same parameters and color coding. To guide the eyes, we have rescaled the $B=m_\pi^2$ case by $1/200$ and the $B=0$ case by $10$.\label{fig_bwo}}
\end{figure}

With all these cares, the solution for $E^U$ has the following expansion near $z=0$,
\be
E^U(z)=E^{(0)}+E^{(2)}z^2+E^hz^2\log z\cdots\,.
\ee
The spectral function $\chi(\omg,k)$ is defined as the imaginary part of the retarded current correlator,
\be
\chi(\omg,k)=-{\rm Im} \frac{(\pi T)^2E^{(2)}}{8\pi G_5E^{(0)}}\,.
\ee
In Figure \ref{fig_spectral} (left figure), we show snapshots of spectral function at different times of thermalization. Typical spatial momentum relevant to heavy ion collisions is $\sim 1$ fm, corresponding to $k\sim 200$ MeV. We restrict ourselves to the region below the lightcone, where we expect to find a chiral magnetic wave peak. We also show the equilibrium thermal spectral function as a reference (right figure).
We discuss the salient features in Figure \ref{fig_spectral}: 

i) The snapshots of spectral functions are taken at times before the break-down of the quasi-static approximation. We observe that a sharp peak appears out of the background plateau, which 
we identify as the out-of-equilibrium chiral magnetic wave.

ii) The peak sits close to the left edge of the plateau. The left edge of the plateau is almost vertical. Further to the left, the spectral function vanishes identically. The location of the left edge can be understood analytically: due to the warping factor $f(z_s)$, there is a mismatch between the frequencies in the upper and lower region $\omg^L=\omg/\sqrt{f(z_s)}$. When $\omg_L$ crosses $k$ from below, the solution in the lower region changes from an exponentially decaying real function to a complex-valued in-falling solution (more specifically, it changes from a modified Bessel function to a Hankel function). The appearance of in-falling wave induces a flux toward IR resulting in a non-vanishing imaginary part of the current correlator. Indeed, we have verified numerically that the location of the left edge is given by $\omg=\sqrt{f(z_s)}k$ with very high accuracy. The right ridge of the plateau in different snapshots seem to lie on top of each other.

iii) We parametrize the location of our chiral magnetic wave peak by
\be\label{peak}
\omg=\sqrt{f(z_s)}k+\Dlt\omg(k,B),
\ee  
where the first piece is the left edge we discussed in (ii) and we have indicated that $\Dlt\omg$ is a function of $k$ and $B$. If we naively extrapolate (\ref{peak}) to the equilibrium limit, i.e. $z_s\to 1$, the first term goes to zero and we hope the second term $\Dlt\omg$ can reproduce the equilibrium chiral magnetic wave. We have studied the dependence of $\Dlt\omg$ on $k$ and $B$, and do find the features characterizing chiral magnetic wave. Figure \ref{width} shows that $\Dlt\omg$ has an excellent linear dependence on $B$ and approximate linear dependence on $k$. These are indeed the behaviors of chiral magnetic wave in the small magnetic field and long wave length limit. However we mention that the precise connection between out-of-equilibrium chiral magnetic wave we found and the one in equilibrium is only suggestive, because quasi-static approximation prevents us from going further in time. For the parameters we explored, the group velocity receives most of its contribution from the first term in (\ref{peak}), which is significantly larger than the group velocity of the equilibrium chiral magnetic wave. This indicates that the out-of-equilibrium chiral magnetic wave moves the chiral charges much faster, potentially enhancing its physical effects in out-of-equilibrium conditions.

iv) Note that the chiral magnetic wave for right-handed charges is expected to move to the same direction as the magnetic field (which means $\Dlt\omg>0$ for $k>0$ and $B>0$). Our result is consistent with this expectation. To check this more clearly, we have calculated the spectral functions with magnetic field reversed ($B<0$) or turned off ($B=0$), while keeping the same spatial momentum: see Figure \ref{fig_bwo} (left figure). We confirm that the peak structure disappears in the $\omg>0$ axis in these cases. We have also calculated the real part of the retarded correlator: see Figure \ref{fig_bwo} (right figure). In the cases of $B<0$ or $B=0$, we see a  structure at $\omg=\sqrt{f(z_s)}k$, which marks the transition between the in-falling wave and the exponentially decaying real function at the left edge we discussed before. In the case of $B>0$, the structure is shifted away from the left edge to the right, in accordance with the analysis of the imaginary part.

v) We have performed our analysis for different values of IR cutoff: $z_c=1/\sqrt{B}, 0.2/\sqrt{B}$ and $2/\sqrt{B}$ to check how sensitive our results are to the IR cutoff. Different values of IR cutoff change the overall normalization of the spectral function, but do not change our results for the peak location which are robust. On the other hand, we have also investigated the case with IR cutoff removed, i.e. the equation (\ref{eom_probe}) with $(6\ka eBz)^2$-term kept all the way through. In this case, the solution in the lower region is always exponentially decaying and is given by
\be
E_L=e^{-3\ka Bz^2}U(\frac{k^2-\omg_L^2}{24\ka B},0,6\ka Bz^2)\,,
\ee
where $U$ is the confluent hypergeometric function. As expected, this case gives vanishing spectral function up to delta-function peaks which are not captured numerically. Therefore, our IR cutoff is crucial for capturing the correct IR physics.

Before we close this section, it is interesting to note how the spectral function develops its non vanishing smooth part in the $\omg<k$ region as the medium thermalizes. Recall that the spectral function is gapped for $\omg<k$ in vacuum. In the thermalizing medium, we observed above that the gap shrinks as the medium thermalizes $z_s\to 1$. This is a manifestation of the mismatch in frequencies $\omg_L=\omg/\sqrt{f(z_s)}$. As the medium thermalizes, $\sqrt{f(z_s)}$ goes to zero, eventually closing the gap. This feature is generic in the gravitational collapse model of thermalization and insensitive to the presence of the magnetic field.


\section{Conclusion}

We have studied chiral magnetic conductivity and chiral magnetic wave in out-of-equilibrium conditions that undergo thermalization. Within the quasi-static approximation, we focused on far out-of-equilibrium region and explored the parameters relevant for RHIC and LHC. For the chiral magnetic conductivity, we considered both its magnitude and the time delay in response. We found that the magnitude is insensitive to the frequency of the magnetic field while the time delay grows with the frequency. This is in contrast to ordinary electric conductivity. As a function of time, both the magnitude and time delay grows, which can be understood as more and more thermalized constituents become available as the system thermalizes.

For the chiral magnetic wave, far away from equilibrium, we found a sharp peak structure in $\omg$ below the lightcone in the spectral function, signaling the out-of-equilibrium chiral magnetic wave. The peak structure is unique to the magnetic field and is a manifestation of anomaly. The location of the peak in $\omg$ relative to a kinematical edge $\sqrt{f(z_s)}k$ depends linearly on the momentum and the magnetic field, which is a feature same to those of the chiral magnetic wave in equilibrium. However, the group velocity 
receives a sizable contribution from the kinematic $\sqrt{f(z_s)}k$ piece which makes the wave moving much faster than in the equilibrium. The correct physics origin of this behavior is not completely clear to us.

The results of this work can be generalized in two aspects: the first is to look at chiral magnetic wave beyond the weak magnetic field limit by including the back reaction of the magnetic field to the metric \cite{D'Hoker:2009mm}. This will be more relevant for LHC, which is expected to produce much stronger magnetic field with only a modest increase of the temperature. The second, perhaps more interestingly, is to go beyond the quasi-static approximation. This can be achieved by raising the temporal resolution and lowering the frequency resolution. This should allow us to extend the coverage of our analysis to near-equilibrium situations and to address the question on the transition from the out-of-equilibrium chiral magnetic wave to the equilibrium chiral magnetic wave.

\vskip 1cm \centerline{\large \bf Acknowledgement} \vskip 0.5cm

We thank Xu-Guang Huang, Dima Kharzeev, Jinfeng Liao, Kiminad Mamo, Larry McLerran, Todd Springer, Misha Stephanov, Derek Teaney, Raju Venugopalan, Yi Yin for useful discussions. SL is supported by RIKEN Foreign Postdoctoral Researchers Program.

\appendix

\section{Alternative calculation of chiral magnetic conductivity}

Let us start by recalling the power expansion in $k$.
\be
A_2(z)=a_2(z)+{\cal O}(k^2)\,,\quad A_3(z)=k a_3(z)+{\cal O}(k^3)\,,
\ee
where $a_2$ and $a_3$ satisfy the equations
\bear
\partial_z\left({f\over z}\partial_z a_2^U\right)+{1\over z}{\omega^2\over f}a_2^U
&=&0\,,\nonumber\\
\partial_z\left({f\over z}\partial_z a_3^U\right)+{1\over z}{\omega^2\over f}a_3^U
-12 i\kappa Q z  a_2^U&=&0\,,\nonumber\\
z\partial_z^2 a_{2}^L-\partial_z a_{2}^L+z\omega_L^2a_{2}^L&=&0\,,\nonumber\\
z\partial_z^2 a_{3}^L-\partial_z a_{3}^L+z\omega_L^2a_{3}^L&=&0\,.
\eear
In the lower region, $a_2^L$ and $a_3^L$ decouple and the solutions are given by
Hankel functions with their ratio unfixed. Requiring the vanishing of $a_3^{(0)}$
would need fine tuning of the ratio. This is actually not needed. Suppose we 
start with a solution in the lower region with arbitrary ratio. Matching it to
the solution above and integrating to the boundary, we obtain the following
electric and magnetic fields to order $k$:
\be\label{eb}
eE_2=-i\omg a_1^{(0)},\quad eE_3=i\omg ka_3^{(0)},\quad eB_2=0,\quad eB_3=ika_1^{(0)},
\ee
and the currents can be extracted from boundary expansion of $a_2^U$ and $a_3^U$:
\be\label{jeb}
J_{EM}^2=\frac{e}{4\pi G_5}a_2^{(2)},\quad J_{EM}^3=\frac{e}{4\pi G_5}ka_3^{(2)}.
\ee
(\ref{eb}) and (\ref{jeb}) are related by electric conductivity $\sig$ and chiral magnetic conductivity $\sig_\chi$:
\be
J_{EM}^2=\sig E_2+\sig_\chi B_2,\quad J_{EM}^3=\sig E_3+\sig_\chi B_3,
\ee
from which we can solve for $\sig$ and $\sig_\chi$ at the same time. It is easy 
to show that the results are independent of the ratio we choose in the lower region.

\section{Region of applicability of quasi-static approximation}

In the quasi-static approximation, we neglect terms proportional to $\dot{z}$
in the continuity condition of $A_\mu\frac{\del x^\mu}{\del \xi^i}$, $F_{\mu\nu}u^\mu\frac{\del x^\mu}{\del \xi^i}$ and $F_{\mu\nu}n^\mu\frac{\del x^\mu}{\del \xi^i}$.
The conditions from $\xi^i=x_2,x_3$ (below we suppress the transverse indices) are given by
\bear
A_x^U&=&A_x^L, \nonumber\\
\del_{t^U}A_x^U\frac{z\dot{z}}{f}+\del_zA_x^U\dot{t^U}zf&=&\del_tA_x^Lz\dot{z}+\del_zA_x^L\dot{t}z, \nonumber\\
\del_{t^U}A_x^U\dot{t^U}+\del_zA_x^U\dot{z}&=&\del_tA_x^L\dot{t}+\del_zA_x^L\dot{z}.
\eear
To neglect the terms proportional to $\dot{z}$ on the left hand side (upper region), we need
\bear\label{ma_above}
\frac{\omg\dot{z}}{f}&\ll&\frac{\del_zA_x^U(\omg)}{A_x^U(\omg)}\dot{t^U}f \,,\nonumber\\
\omg\dot{t}^U&\gg&\frac{\del_zA_x^U(\omg)}{A_x^U(\omg)}\dot{z}.
\eear
Using (\ref{june1}) and (\ref{june2}), we obtain
\bear
\frac{\omg\dot{z}}{\sqrt{f}}&\ll&\frac{\del_zA_x^L(\omg/\sqrt{f})}{A_x^L(\omg/\sqrt{f})}\dot{t}^Uf, \nonumber\\
\frac{\omg\dot{t}^Uf}{\sqrt{f}}&\gg&\frac{\del_zA_x^L(\omg/\sqrt{f})}{A_x^L(\omg/\sqrt{f})}\dot{z}.
\eear
Similarly, to neglect the terms proportional to $\dot{z}$ on the right hand side (lower region), we obtain
\bear\label{ma_below}
\frac{\omg\dot{z}}{\sqrt{f}}&\ll&\frac{\del_zA_x^L(\omg/\sqrt{f})}{A_x^L(\omg/\sqrt{f})}\dot{t}^Lf, \nonumber\\
\frac{\omg\dot{t}^L}{\sqrt{f}}&\gg&\frac{\del_zA_x^L(\omg/\sqrt{f})}{A_x^L(\omg/\sqrt{f})}\dot{z}.
\eear
Obviously (\ref{ma_below}) is included in (\ref{ma_above}). Using solution of $A_x^L$ in terms of Hankel function, we end up with
\bear
\dot{z}&\ll& \frac{H_0^{(1)}(\omg z/\sqrt{f})}{H_1^{(1)}(\omg z/\sqrt{f})}\sqrt{f+\dot{z}^2} \,,\nonumber\\
\sqrt{f+\dot{z}^2}&\gg&\frac{H_0^{(1)}(\omg z/\sqrt{f})}{H_1^{(1)}(\omg z/\sqrt{f})}\dot{z}.
\eear

 \vfil

\end{document}